\documentclass[12pt]{article} 
\usepackage{ccaption}
\usepackage{amssymb}
\usepackage{times}
\usepackage{titling}
\usepackage{graphicx}
\usepackage{amsmath}
\usepackage[version=3]{mhchem}
\usepackage{color}
\usepackage{bm}
\usepackage{lineno}
\usepackage{multirow}
\usepackage{float}
\usepackage{scicite}

\usepackage[paper=letterpaper,twoside,top=2.0cm,bottom=2cm,left=1.5cm,         right=1.5cm,bindingoffset=0.0cm,voffset=0.0cm]{geometry}
\usepackage[labelfont=bf,labelsep=period]{caption}
\usepackage[symbol]{footmisc}
\usepackage[pagebackref=true]{hyperref}
\hypersetup{colorlinks=true,linkcolor=blue,urlcolor=blue}
\usepackage[doipre={DOI: }]{uri}

\renewcommand{\refname}{\subsection*{References}}

\newcommand{\blue}[1]{{\color{blue} #1}}

\def\bQ{\bold{Q}}

\renewcommand{\thesection}{}
\renewcommand{\thesubsection}{\arabic{subsection}}  

\def\NiTL{{Na$_2$BaNi(PO$_4$)$_2$}}{}
\title{
{\bf Universal dynamics of a pair condensate}
}
\author{
Qing Huang$^{1,2}$,\, Hao Zhang$^{3}$, Yiqing Hao$^{4}$,\\ Weiliang Yao$^{2}$, Daniel M. Pajerowski$^{4}$,
Adam A. Aczel$^{4}$,\\  Eun Sang Choi$^{5}$, Kipton Barros$^{3}$, Bruce Normand$^{6,7}$,\\
Haidong Zhou$^{2,}$\thanks{Email: hzhou10@utk.edu}, Andreas M. L\"auchli$^{6,7,}$\thanks{Email: andreas.laeuchli@psi.ch}, Xiaojian Bai$^{1,}$\thanks{Email: xbai@lsu.edu}, Shang-Shun Zhang$^{2,}$\thanks{Email: szhang57@utk.edu}\\
\\
\normalsize{$^1$Department of Physics and Astronomy, Louisiana State University, Baton Rouge, LA 70803, USA} 
\\
\normalsize{$^2$Department of Physics and Astronomy, University of Tennessee, Knoxville, TN 37996, USA}
\\
\normalsize{$^3$Theoretical Division and CNLS, Los Alamos National Laboratory, Los Alamos, NM 87545, USA}
\\
\normalsize{$^4$Neutron Scattering Division, Oak Ridge National Laboratory, Oak Ridge, TN 37831, USA}
\\
\normalsize{$^5$National High Magnetic Field Laboratory, Florida State University, Tallahassee, Florida 32310-3706, USA}
\\
\normalsize{$^6$PSI Center for Scientific Computing, Theory and Data, CH-5232 Villigen-PSI, Switzerland}
\\
\normalsize{$^7$Institute of Physics, Ecole Polytechnique F\'ed\'erale de Lausanne (EPFL), CH-1015 Lausanne, Switzerland}
} 

\date{\today}

\begin{document}

\maketitle


\subsection*{Abstract} 
\baselineskip20pt
{\bf Pair condensates appear in multiple branches of physics, always introducing exotic phenomena. The pair condensate in quantum magnetism is the spin nematic, whose static (quadrupolar) order is difficult to access, favoring dynamical probes. Here, we perform high-resolution neutron spectroscopy to obtain direct evidence for the presence of two spin-nematic phases induced in the triangular-lattice antiferromagnet \NiTL~by controlling the applied magnetic field. By combining precise experiments with quantitative theoretical and numerical analysis, we identify universal dynamics arising from the pair condensate. We show explicitly how the gapless Goldstone mode influences the dispersion and induces Cherenkov-like velocity-selective decay of the gapped single-quasiparticle band. These common spectral features shed new light on spin-nematic dynamics and underline the universal phenomenology shared by pair condensates across different physical systems.}


\subsection*{Main Text}

The pairing of elementary quasiparticles adds a completely new ingredient to the behavior of quantum matter. Because bound pairs of identical fermions or bosons are necessarily bosonic, at low temperatures they can undergo Bose-Einstein condensation (BEC) into a macroscopic quantum phase known as a pair condensate, from which a variety of extraordinary phenomena emerge. Prominent examples of fermionic pair condensates include $^3$He~\cite{lee1997} and ultra-cold atoms in fermionic superfluids~\cite{anderson1995observation,bradley1995evidence, davis1995bose}, Cooper pairs of electrons in superconductors~\cite{cooper1956bound, bardeen1957theory}, and electron-hole condensation in the excitonic ordered phase of narrow-gap semiconductors~\cite{jerome1967,Kohn1970,kaneko2025}. Although less common because bosons can condense singly, bosonic pair condensation also takes place with ultra-cold atoms in the ``molecular BEC''~\cite{zhang2021} and with bound magnon pairs in the spin-nematic phase of quantum magnets~\cite{orlova2017,kohama2018,fogh2024b}. While the physics of these condensates may vary widely in length scale, energy scale, and type of observable, one common property is the breaking of a continuous ${\mathrm U}(1)$ symmetry, which is generally that associated with particle number conservation. This symmetry-breaking leads to distinctive signatures in the excitation spectrum~\cite{stewart2008using,schirotzek2008,feld2011observation, hoinka2017goldstone,biss2022excitation}, which are universal to pair condensates across all their different realizations.

\begin{figure}[h!] 
 	\begin{center}
 		\includegraphics[width=1\textwidth]{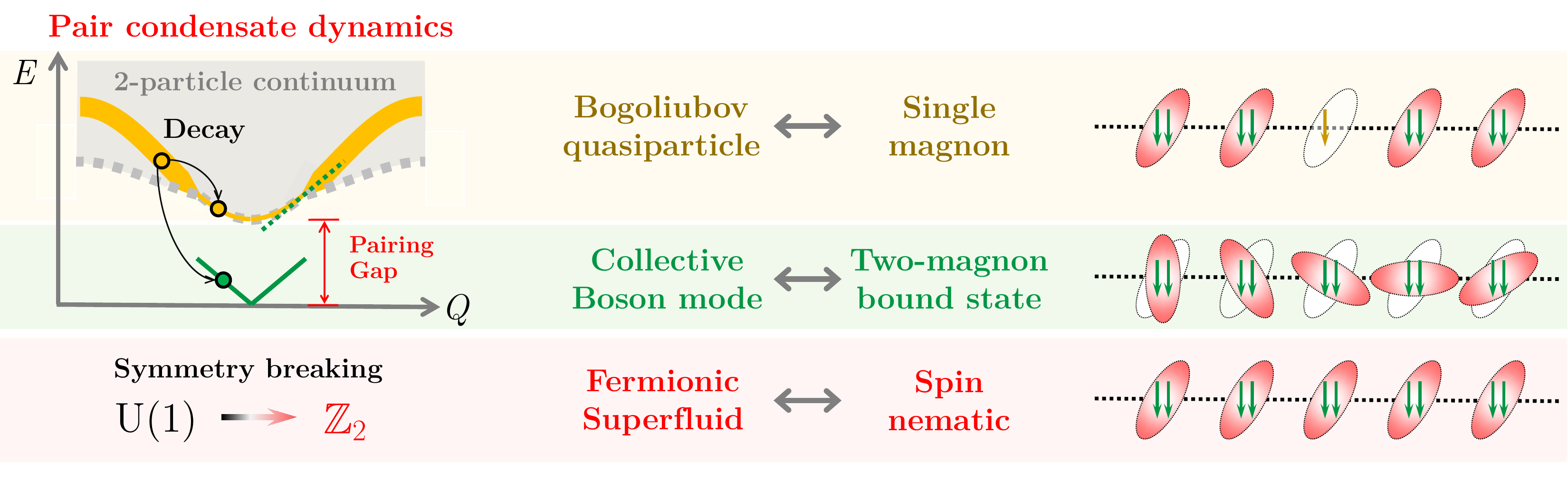}
	\end{center}
 	\caption{\label{fig1}
{\bf Universal dynamics of a pair condensate.} Representation of the physics common to pair condensates ranging from the fermionic superfluid to the spin nematic. Single-particle excitations have finite energies (yellow band). By forming pairs, they profit from the binding energy (pairing gap) to condense into the ground state. Collective bosonic pair excitations form the gapless Goldstone mode (green band). Its existence allows dynamical decay processes of the single-particle excitations that involve the emission of a Goldstone mode (black arrows), which form a two-particle continuum (gray area) in the spectrum. Near the single-particle band-bottom is a decay-free region that extends to the point where the velocity is equal to that of the Goldstone mode (dashed green line). 
Slower quasiparticles have no decay processes, while faster quasiparticles create the analog of Cherenkov radiation, decaying as they push through the pair-condensate medium at ``superluminal'' velocity. At right we represent how the condensate and its excitations are realized in quantum magnets. Single magnons (yellow arrow) are the high-energy bosonic quasiparticles. Two-magnon bound states (TMBS, double green arrows) gain pairing energy and condense into the ground state. This state is the spin nematic, whose directional order parameter is represented by the red ellipsoids. The Goldstone mode is a collective phase modulation of the nematic order.}
\end{figure}

The characteristic excitations of a generic pair condensate, represented in Fig.~\ref{fig1}, are a gapped, high-energy single-particle band and a gapless Goldstone mode at low energies. The energetically favored bound pair states condense into the ground state and their collective excitations are gapless, while the single-particle spectrum retains a finite gap given by the pair binding energy. This gapped, ``pair-breaking'' single-particle mode, familiar from the Bogoliubov quasiparticles of a superfluid~\cite{leggett1975} or superconductor~\cite{bardeen1957theory}, is a universal hallmark of all pair condensates. However, in some pair condensates the Goldstone mode can acquire a gap due to additional interactions, such as the Anderson-Higgs mechanism in superconductors \cite{anderson1963} or electron-phonon couplings in excitonic insulators \cite{Murakami2020}. Thus our focus here is on the generic case of electrically charge-neutral pair condensates, which possess additional universal dynamical properties due to the presence of the gapless Goldstone mode. 

Specifically, the breaking of ${\mathrm U}(1)$ symmetry makes possible a decay process in which a single particle loses energy by emitting a Goldstone mode. However, this can occur only when its velocity exceeds that of the Goldstone mode, which as Fig.~\ref{fig1} shows is not the case around the band minima of a gapped, quadratic and a gapless linear mode. Single-particle excitations are then stable over a small region of momentum and energy determined by the Goldstone-mode dispersion, but outside this window they show a characteristic decay, creating a key dynamical fingerprint in the line width. In a broader context, this interplay between a massive and a massless particle is analogous to Cherenkov radiation~\cite{cerenkov1934}, where a charged particle travelling through a dielectric medium emits light when its speed exceeds the phase velocity of light in that medium. 

Quantum magnetic insulators present a particularly suitable platform for the systematic study of condensation phenomena, including controlled BEC and different types of pair condensate. Magnetic quasiparticles (magnons) are bosons that, depending on their mutual interactions, can undergo field-induced BEC~\cite{Giamarchi2008,Zapf2014} or can pair into two-magnon bound states (TMBS)~\cite{Wortis1963,Hanus1963,silberglitt1970effect,ono1971two,oguchi1971theory}. The TMBS pair condensate is called the spin nematic~\cite{andreev1984spin, zhitomirsky2010magnon}, in direct reference to its ${\mathbb Z}_2$ symmetry, and has magnetic quadrupolar order without dipolar order. Predicted in theoretical models for many years~\cite{blume1969biquadratic, Andreas2006, shannon2006nematic, wierschem2012magnetic}, the spin nematic remains one of the more elusive states of quantum matter, with experimental reports emerging only recently~\cite{orlova2017,kohama2018,kim2024quantum,fogh2024b}. Given the lack of experimental probes that couple directly to quadrupole moments, spectroscopic measurements provide a compelling alternative for identifying spin-nematic phases~\cite{fogh2024b}. A particular advantage of quantum magnets is that inelastic neutron scattering (INS) provides high-resolution spectra over the full momentum space, allowing accurate quantitative modelling that reveals valuable insight into the emergence of universal dynamics.

Here, we demonstrate that the key dynamical signatures of a pair condensate, namely the massive single-particle mode, the massless Goldstone modes, and their mutual  ``Cherenkov radiation'' effect, can be observed in the quantum magnet \NiTL. By controlling the applied magnetic field, we sweep the system through all four magnetic phases predicted by theory, fully polarized (FP), ferroquadrupolar (FQ), collinear up-up-down three-site order (UUD), and at zero field a nematic supersolid (NSS). The FP and UUD phases have U(1) spin symmetry, whereas the FQ and NSS phases are both spin nematics with ${\mathbb Z}_2$ symmetry, presenting two pair-condensation transitions. In all four phases we use INS to measure the magnetic spectra at the highest available resolution, revealing their nature through their distinctive quasiparticle dispersion and decay. We use a generalized spin-wave theory (GSWT) to explain how the quasiparticle dynamics are driven by the underlying spin-nematic order and provide quantitative validation by comparing the INS data with exact diagonalization (ED) calculations of the full many-body response. These results constitute a clear confirmation of universal behavior in a pair condensate and provide a detailed spectral characterization of elusive spin-nematic phases in frustrated quantum magnetic materials.

\begin{figure}[h!] 
 	\begin{center}
 		\includegraphics[width=1\textwidth]{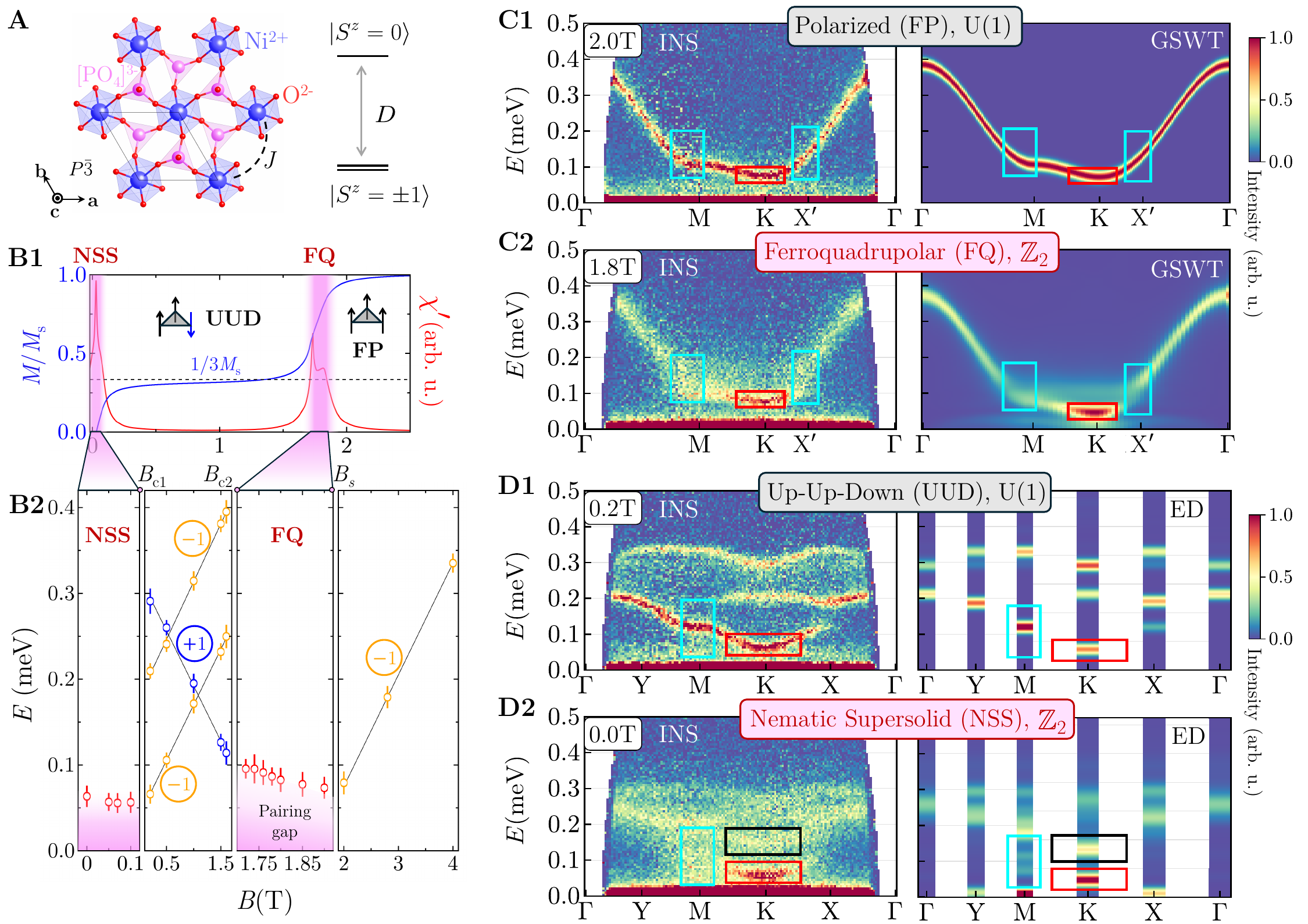}
	\end{center}
 	\caption{\label{fig2}
{\bf Spectroscopic signatures of two quantum spin-nematic phases in \NiTL.} ({\bf A}) Crystal structure of a single triangular layer of Ni$^{2+}$ ions. The $S = 1$ spins are subject to a nearest-neighbor superexchange interaction $J$ and an easy-axis single-ion anisotropy $D$; in \NiTL, $D \approx 4J$ is dominant. ({\bf B}) Experimental phase diagram obtained from bulk a.c.~susceptibility ($\chi^\prime$) measurements (B1) and INS measurements of the single-magnon energy at the K point (B2). Both quantities were measured at $T = 0.05$~K with $\textbf{B} \parallel \textbf{c}$. The magnetization ($M$) was obtained by integrating the a.c.~susceptibility data. Circled numbers denote the magnon spin quantum numbers $\Delta S^z$. 
({\bf C})({\bf D}) Spin excitation spectra of the four field-induced quantum phases in \NiTL. The fully polarized (FP) and up-up-down (UUD) phases are U(1)-symmetric; the ferroquadrupolar (FQ) and nematic supersolid (NSS) phases are spin nematics with ${\mathbb Z}_2$ symmetry. High-resolution INS data along a high-symmetry path in momentum space (left) is shown with corresponding theoretical calculations (right) performed by $\mathrm{SU}$(3) GSWT (C) and by ED (D). 
Turquoise boxes highlight the areas where significant energy broadening occurs, reflecting strong magnon decay. Red boxes highlight the decay-free region at the bottom of the single-particle band where the mode remains sharp. Black boxes in the NSS phase highlight an additional excitation not present in the UUD phase.}
\end{figure}

\NiTL~crystallizes in the trigonal space group P$\bar{3}$, hosting perfect triangular-lattice planes of Ni$^{2+}$ ions in an octahedrally-coordinated environment, as shown in Fig.~\ref{fig2}A. A trigonal distortion of the oxygen octahedra results in an easy-axis anisotropy $D \approx 1.6$ K that splits the local effective $S = 1$ manifold [Fig.~\ref{fig2}A]. The phosphate anions form superexchange pathways between the Ni$^{2+}$ ions that create a nearest-neighbor interaction $J$ between the magnetic moments; although $J$ is weak compared to $D$ ($\approx 4J$), this small energy scale makes it possible to scan the entire phase diagram using moderate magnetic fields. However, it also makes the ground-state properties and low-lying modes difficult to resolve, and previous studies \cite{ding2021, li2021a, sheng2025bose} have inferred the presence of the FP, FQ, UUD, and NSS phases from thermodynamic measurements at low temperatures and from the field-induced reduction of the TMBS gap in the FP phase. While the FP and FQ phases have the translational symmetry of the triangular lattice, the UUD phase has $1/3$ of the saturation magnetization [the plateau phase in Fig.~\ref{fig2}B1] and a three-sublattice structure, a spin-density modulation common to the NSS phase. The U(1) spin symmetry of the FP and UUD phases is broken down to ${\mathbb Z}_2$ in the FQ and NSS phases at two continuous phase transitions, whereas the FQ-UUD transition is first-order. The two candidate spin-nematic phases then exist within narrow field ranges just below saturation and just above zero field, as shown in Fig.~\ref{fig2}B. Our goal is to use the observable spin excitations of both these phases for an unambiguous demonstration of their existence and universal properties.


As the first step towards this goal, small, plate-like single crystals of \NiTL~were grown from an NaCl flux, as summarized in the Methods section. Next, multiple single crystals with a total mass of approximately 5~g were coaligned into a crystal mount. We then performed two sets of INS measurements on the CNCS spectrometer at the Spallation Neutron Source (ORNL) at a base temperature of 50~mK, using a very low incident energy of $E_\text{i} = 1$~meV to achieve an energy resolution of 0.025~meV (FWHM) at the elastic line (Methods section and Sec.~S2.1 of the Supplementary Materials (SM) \cite{sm}). In the FP and UUD phases, where ${\mathrm U}(1)$ spin rotation about the easy axis is an exact symmetry, the excitations consist of well-defined single-magnon modes, whose line widths are resolution-limited, as shown at $B = 2.0$~T in Fig.~\ref{fig2}C1 and at $0.2$~T in Fig.~\ref{fig2}D1. The number of magnon modes, given by the number of sites in the magnetic unit cell, triples in the UUD phase. All of these magnons display a rigid, linear shift with the applied field, consistent with the Zeeman energy $\Delta E_\text{Zeeman} = - g \mu_\text{B} \Delta S^z B$ (their gaps are shown by the yellow and blue symbols in Fig.~\ref{fig2}{B}2), where the spin quantum numbers $\Delta S^z =\pm  1$ are conserved. However, this linear behavior is disrupted in the putative spin-nematic regimes [red symbols in Fig.~\ref{fig2}{B}2], halting the closure of the magnon gaps and indicating a transition of the ground state into a different spin sector.

\begin{figure}[h!] 
 	\begin{center}
 		\includegraphics[width=1.\textwidth]{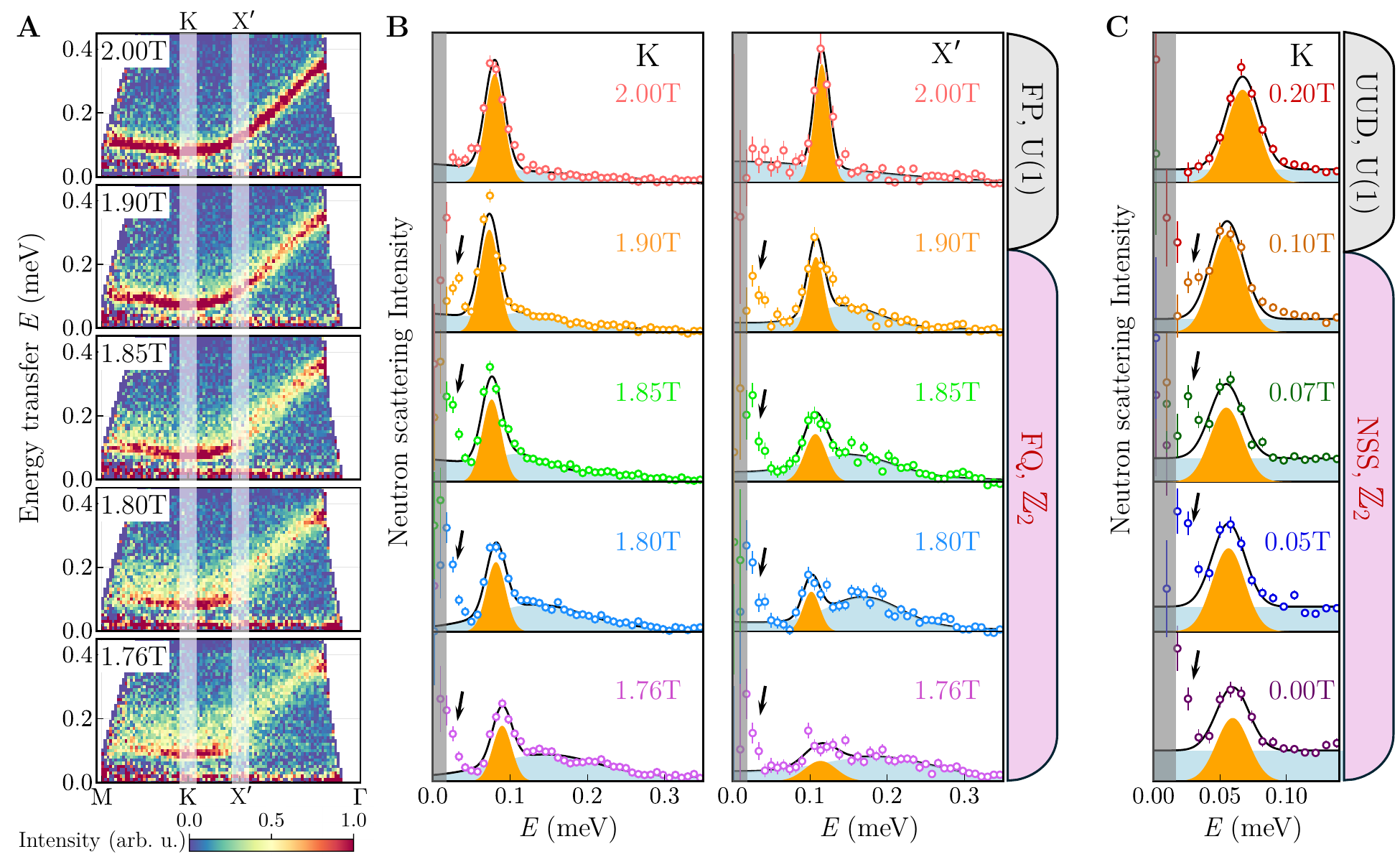}
	\end{center}
 	\caption{\label{fig3}
{\bf Excitation spectra of quantum spin-nematic phases in \NiTL.} ({\bf A}) Field-induced evolution of the background-subtracted INS spectrum across the FP-FQ phase transition, showing systematic broadening of the single-magnon bands toward higher energies as the field is decreased through the FQ phase. The background to subtract was determined from a measurement performed deep in the FP phase, at 5 T (Methods section). ({\bf B}) Analysis of constant-{\bf Q} spectra taken at the K and X$^\prime$ points, using the data integration ranges marked by white stripes in panel (A). Black curves represent full Gaussian peak fits to a single-magnon branch (orange) and a broad scattering continuum (light blue shading). Black arrows highlight the emergence of an additional scattering contribution at very low energies. ({\bf C}) Field-induced evolution of the spin excitations across the UUD-NSS phase transition, showing the strong resemblance between the low-energy spectral features of the NSS and FQ phases. The fitted values of the single-magnon gaps (i.e.~at K) in both phases are shown as the red points in Fig.~2B2. Gray shaded areas mark the region masked by the elastic line.}
\end{figure}

By contrast, the single-magnon spectrum develops a continuum-like upward broadening in the FQ phase, at $B = 1.8$~T in Fig.~\ref{fig2}C2, and three highly structured continua in the NSS region at 0~T in Fig.~\ref{fig2}D2. The bottom edges of the low-lying continua, highlighted by red boxes, retain the strong intensities seen respectively in the FP and UUD phases, but the spectra outside this region (examples are indicated by blue boxes) become substantially less intense due to their strong broadening. To investigate this behavior, we tracked the spectral  evolution at five selected fields across each of the spin-nematic phase transitions, as shown in Fig.~\ref{fig3}. Focusing on the FP-FQ transition, we observe at the K point (inside the red box) that the single-magnon peak of the FP phase remains close to the same energy [orange shaded areas in Fig.~\ref{fig3}B], with a resolution-limited width $w \simeq 0.025$~meV (FWHM) throughout the entire field range. It remains the dominant spectral feature, while a tail of intensities develops gradually at higher energies. In contrast, at the X$^\prime$ point (${\bf Q} = (2/9,2/9,0)$, inside a blue box) the magnon intensity fades rapidly as the field decreases, transferring its spectral weight into a broad continuum (light blue shading). These observations verify the persistent single-particle gap in a pair condensate.


The next essential dynamical signature is the Goldstone mode that emerges as the ${\mathrm U}(1)$ symmetry is broken. The Goldstone-mode intensity vanishes as its energy approaches the $\Gamma$ point with a linear slope \cite{Smerald2015}, but it has finite spectral weight at other momenta. To search for Goldstone modes at very low energies in the FQ and NSS phases, we subtracted a background measurement collected at $B = 5$~T, where the single magnon lies at much higher energy, effectively removing the strong elastic signal. A clear residual intensity is observed consistently below the single-magnon gap in both the FQ and NSS phases, as indicated by the black arrows in Figs.~\ref{fig3}B and \ref{fig3}C, which is not present in the FP and UUD phases (top panels). Additionally, although not directly observable, the threshold velocity of the Cherenkov-like decay seen in the single-magnon mode effectively quantifies the velocity of the Goldstone mode. While these data provide strong evidence for the presence of Goldstone modes in both spin nematics, experiments with higher energy resolution would be necessary to map their full dispersions. 


Turning now from our experimental data to a theoretical description, the minimal model required to explain the spin-nematic phenomenology in \NiTL{} is the $S = 1$ antiferromagnet on the triangular lattice, 
\begin{equation}\label{eq:model}
{\cal H} = J \sum_{\langle ij \rangle} \left( S_{i}^{x}S_{j}^{x} + S_{i}^{y}S_{j}^{y} + \Delta S_{i}^{z}S_{j}^{z} \right) - D \sum_{i} \left( S_{i}^{z} \right)^{2} - g \mu_{B} B \sum_{i} S_{i}^{z},
\end{equation}
with $J$ and $D$ as above, $\Delta$ a symmetry-allowed XXZ anisotropy, and $B$ the applied field. Here $\langle ij \rangle$ denotes nearest-neighbor sites only and 
$S^{\mu}_i$ ($\mu = x,y,z$) are the spin-$1$ operators. Although an interlayer coupling induces three-dimensional magnetic order at low temperatures, it is too weak to cause any discernible dispersion in the out-of-plane direction (Sec.~S2.1 of the SM \cite{sm}). The four field-induced ground states of this model in the easy-axis limit have been studied in theory~\cite{Wessel2005, Heidarian2005, Melko2005, Boninsegni2005, moreno2014case,Savary2022} and recently in experiment~\cite{sheng2025bose}. To extract the model parameters from our spectroscopic data, we performed a combined fit of the single-magnon modes measured in the FP phase to calculations performed by ${\mathrm{SU}}(3)$ GSWT (Methods section and Sec.~S2.2 of the SM \cite{sm}) and in the UUD phase to ED calculations (Methods section and Sec.~S2.3 of the SM \cite{sm}). The best fits in these two phases, shown respectively in the right panels of Figs.~\ref{fig2}C1 and \ref{fig2}D1, yield the optimal parameters $J = 0.0348(2)$ meV, $D = 0.138(1)$ meV, and $\Delta = 1.06(1)$, with $g = 2.245(6)$ (full details are presented in Sec.~S2.5 of the SM \cite{sm}). These parameters represent a refinement of the results deduced in Ref.~\cite{sheng2025bose}. Next, we provide a qualitative understanding of the observed dynamics based on symmetry arguments, followed by a summary of our quantitative calculations.


By conservation of angular momentum, INS detects only transverse excitations, whose spin quantum numbers differ from the ground state by $\Delta S^z = \pm 1$, and longitudinal spin excitations with $\Delta S^z = 0$. In the ${\mathrm U}(1)$-symmetric phases (FP and UUD), the single-magnon excitations have $\Delta S^z = \pm 1$, and hence are detectable, whereas the TMBS, with $\Delta S^z = \pm 2$, is not. The breaking of ${\mathrm U}(1)$ symmetry in the spin-nematic phases (FQ and NSS) means that $S^z$ is no longer conserved, but the residual ${\mathbb Z}_2$ symmetry makes the spin parity, ${\rm mod}(S^z, 2)$, the appropriate quantum number for characterizing different eigenstates. The pair-condensate ground states are parity-even, the single-magnon excitations are parity-odd, and the TMBSs are parity-even, making them visible in INS as longitudinal fluctuations.

Symmetry considerations also strongly constrain the single-magnon decay processes responsible for the spectral broadening observed in our INS data. Because all interaction terms in Eq.~\eqref{eq:model} preserve ${\mathrm U}(1)$, the initial and final states involved in the decay process must share the same spin quantum number. In the FP phase, the single magnon and TMBS carry well-defined spin quantum numbers, respectively $-1$ and $-2$, and hence a single magnon cannot decay into any multi-particle states formed from magnons and TMBSs. In the FQ phase, however, all excitations are superpositions of states with odd and even quantum numbers, and in general any decay processes preserving spin parity are allowed. Thus the decay of a single magnon into two magnons is forbidden, whereas decay into another single magnon and a TMBS is allowed. The same considerations apply to the UUD and NSS phases, subject to the kinematic constraints of energy and momentum conservation.

\begin{figure}[h!] 
 	\begin{center}
 		\includegraphics[width=0.65\textwidth]{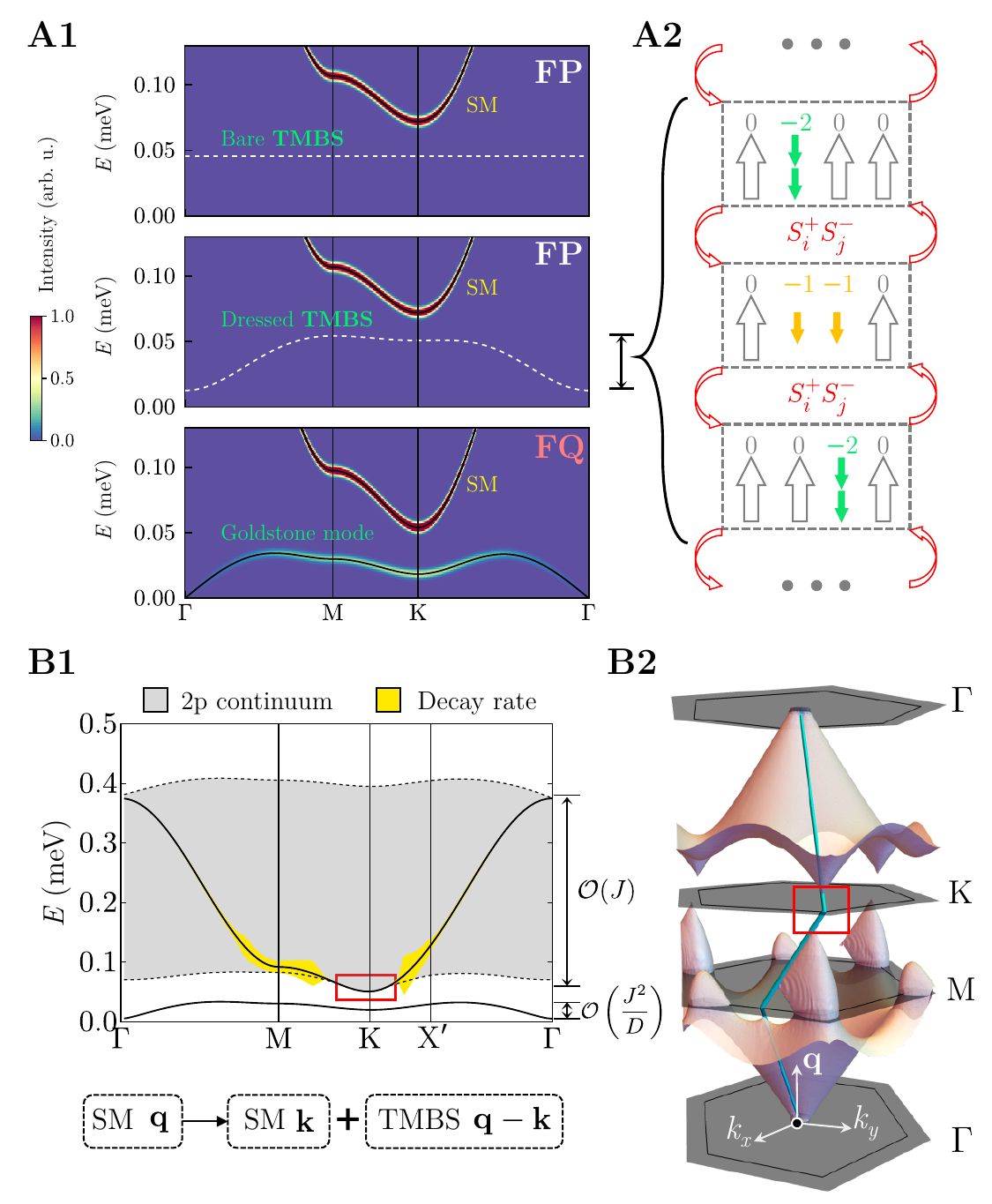}
	\end{center}
 	\caption{\label{fig4}
{\bf Ferroquadrupolar transition and quantum decay dynamics.} ({\bf A1}) Visualization of dynamics at the FQ-FP transition due to TMBS mobility. Top: an immobile (localized) TMBS has a flat energy band . Middle: weak TMBS hopping results in a weakly dispersive band. Bottom: lowering the magnetic field causes pair condensation, transforming the TMBS band into the Goldstone mode. ({\bf A2}) Higher-order quantum processes responsible for TMBS mobility. The change in spin quantum number ($\Delta S^z$) is 0 for the ground state (empty gray arrows), $-1$ for a single magnon (solid yellow arrows), and $-2$ for a TMBS (double green arrows).
({\bf B1}) Kinematic conditions for single-magnon decay in the FQ phase. A single magnon (SM) at wave vector {\bf q} decays into another at {\bf k} by emitting a TMBS at $\mathbf{q} - \mathbf{k}$. The upper solid black curve depicts the single-magnon dispersion, $\omega_{\bf q}$, and the lower the TMBS dispersion, $\varepsilon_{\bf q}$. The gray area represents the energy, $\omega_{\bf k} + \varepsilon_{{\bf q}-{\bf k}}$, of the two-particle (2p) continuum formed in the decay process, which takes place when $\omega_{\bf k} + \varepsilon_{{\bf q}-{\bf k}} = \omega_{\bf q}$. 
({\bf B2}) Representation of allowed momenta {\bf k} (pink surface) in the first Brillouin zone at which a decay process is possible when {\bf q} lies on the high-symmetry path (top to bottom). The red line indicates the situation ${\mathbf{q}} = {\mathbf{k}}$. The crucial feature is that the density of decay states vanishes for momenta around K, corresponding to the red box in panel (B1), where the single-magnon band coincides with the lower bound of the two-particle continuum. A finite magnon decay rate is possible when {\bf q} lies outside the red box, as represented by the width of the yellow area.}
\end{figure}

To describe the quasiparticle dynamics, we build these considerations into the ${\mathrm{SU}}(3)$ GSWT framework using perturbation theory. At lowest order, a TMBS is localized on one lattice site, producing a flat band in momentum space [Fig.~\ref{fig4}A1]. At next order, quantum processes allow the TMBS to propagate [Fig.~\ref{fig4}A2], gaining kinetic energy [Fig.~\ref{fig4}A1]. Introducing a TMBS hopping term in Eq.~\eqref{eq:model}, as summarized in the Methods section, leads to a new effective Hamiltonian in which the TMBS is a dispersive band with its global minimum at $\Gamma$. As the field is decreased, the gap of this band closes continuously until the TMBS condenses, inducing a quadrupole moment in the ground state, which is the spin-nematic FQ phase. The TMBS band becomes the gapless Goldstone mode of the pair condensate [Fig.~\ref{fig4}A1] and is observable by INS in the longitudinal spin channel. 


We performed quantitative calculations of the INS intensity in the FQ phase using GSWT for the renormalized Hamiltonian (full technical details are contained in Sec.~S2.2 of the SM \cite{sm}). Single-magnon decay by TMBS emission [Fig.~\ref{fig4}B] is captured at one-loop order and possesses two universal features. First, because the Goldstone-mode velocity (at $\Gamma$) exceeds that of the magnon at its band minimum [Figs.~\ref{fig1} and \ref{fig4}B1], the lower edge of the two-particle continuum coincides with the single-magnon dispersion over a finite region around the K point; this results in zero phase-space volume for the decay of magnons with {\bf Q} close to K, and hence a vanishing decay rate within the red boxes in Fig.~\ref{fig4}B. The energy scale of this decay-free region is comparable to the bandwidth of the Goldstone mode, ${\cal O} (J^2/D)$, which is relatively narrow for large $D/J$ [Fig.~\ref{fig4}B1]. The second feature is the onset of decay once the single-magnon velocity exceeds that of the Goldstone mode, in analogy with Cherenkov radiation. This allows the two-particle continuum to surround the single-magnon band [Fig.~\ref{fig4}B1], providing in a finite phase space for decay [Fig.~\ref{fig4}B2]. The decay-allowed region extends over a wide energy range [${\cal O} (J)$] and the decay rate exhibits a non-universal {\bf Q}-dependence, shown by the yellow shading in Fig.~\ref{fig4}B1, that depends on the density of states in the two-particle continuum and is described quantitatively by GSWT, as we show in Sec.~S2.2 of the SM \cite{sm}. Finally, the quasiparticle weight lost from the single magnon is transferred to the two-particle continuum, forming the broad, flat structure observed in the INS data, meaning that all of the decay-allowed and decay-free properties measured in the single-magnon spectrum of the FQ phase are well reproduced [Fig.~\ref{fig2}C2]. We remark here that the narrow [${\cal O}(J^2/D)$] TMBS bandwidth amplifies quantum fluctuation effects, resulting in a significant single-magnon decay rate that increases with the ratio $D/J$, which highlights again the significance of flat and narrow bands in many-body physics. 

\begin{figure}[h!] 
 	\begin{center}
 		\includegraphics[width=0.7\textwidth]{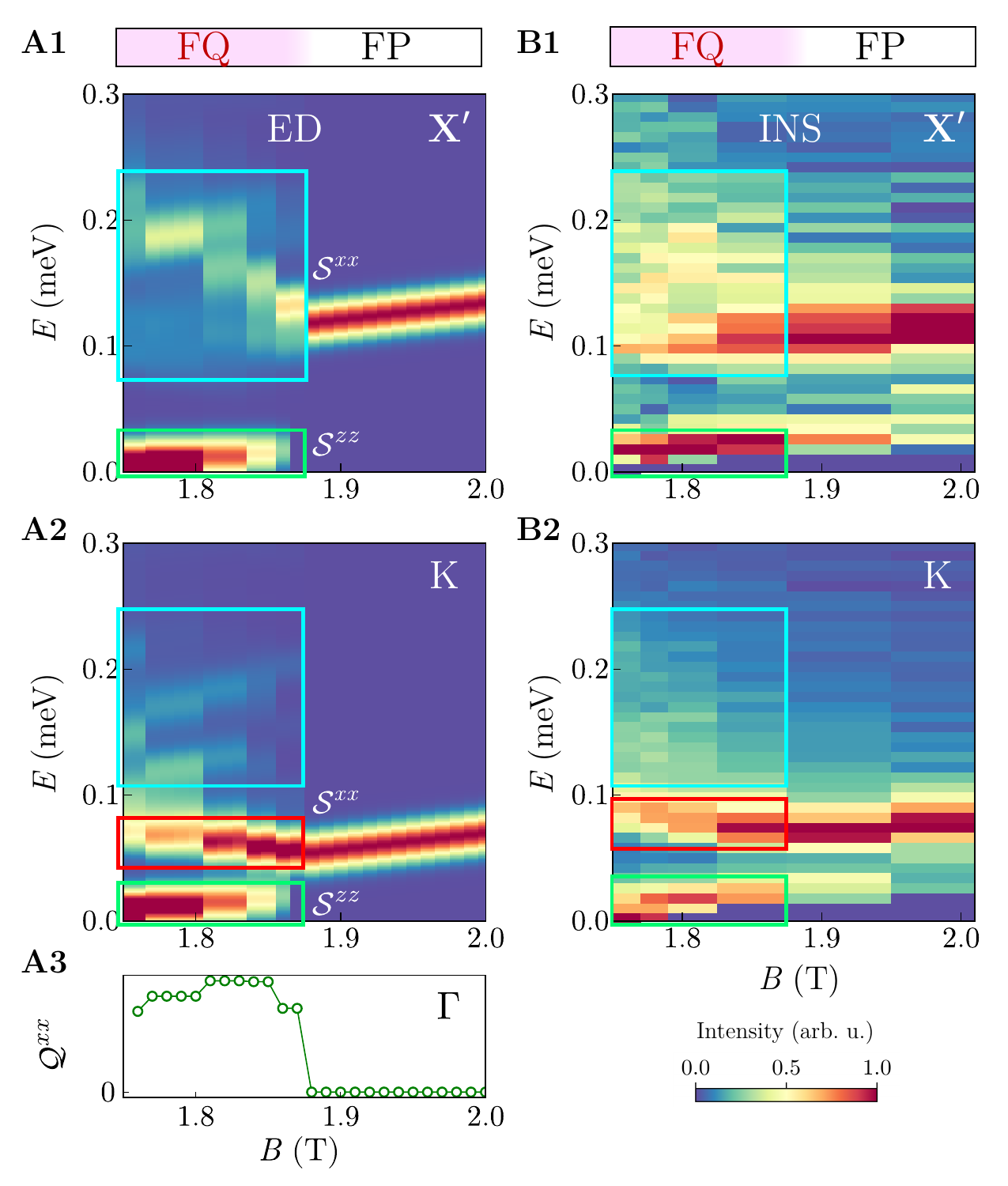}
	\end{center}
 	\caption{\label{fig5}
{\bf Quantitative analysis of excitation spectra through the FQ-FP transition.} ({\bf A1})({\bf A2}) Dynamical spin structure factors at the X$^\prime$ and K points computed by ED on a cluster of 27 spins, separating the transverse ($\mathcal{S}^{xx}$) and longitudinal ($\mathcal{S}^{zz}$) channels. 
({\bf A3}) Quadrupole-quadrupole correlations at the $\Gamma$ point, computed for $N = 27$ and shown as a function of field to confirm that the dramatic spectral changes are caused by the onset of the FQ phase. ({\bf B1})({\bf B2}) Field-dependence of the background-subtracted INS intensities measured at the X$^\prime$ and K points. Green boxes highlight the Goldstone mode, red boxes the bottom of the single-magnon band, and blue boxes the scattering continua that develop in the FQ phase.}
\end{figure}

To benchmark our perturbative calculations and to gain unbiased insight into the dynamical features intrinsic to Eq.~\eqref{eq:model}, we performed ED calculations of the excitation spectra at all fields on cluster sizes up to $N = 36$, as summarized in the Methods section. By their nature, ED spectral data are restricted to small numbers of {\bf Q} values but are not quantized in energy, and the number, energetic position, and spectral weight of the computed poles reflect clearly the presence of a sharp magnon mode or an excitation continuum (Sec.~S2.3 of the SM \cite{sm}). In Fig.~\ref{fig5} we focus on the FP-FQ transition and also separate the transverse ($\mathcal{S}^{xx} + \mathcal{S}^{yy}$) and longitudinal ($\mathcal{S}^{zz}$) spin channels for two momentum transfers representing the red (K) and turquoise (X$^\prime$) boxes in Fig.~\ref{fig2} (a more complete comparison of INS and ED data is presented in Sec.~S2.4 of the SM \cite{sm}). First, the spin-nematic phase transition is very clear in our ED data, while its broader nature in the INS data presumably reflects the finite temperature of the experiment, as we discuss below. Next, at K we find an intense single-magnon band in $\mathcal{S}^{xx}$ (red boxes) with only a very weak scattering continuum at high energies (turquoise boxes), and the Goldstone mode in $\mathcal{S}^{zz}$ at very low energies (green boxes). By contrast, while the Goldstone mode is present in the FQ phase at X$^\prime$, the single magnon disappears quickly in favor of a broad continuum. Quantitative details such as the slight increase of the single-magnon gap as the field is decreased through the FQ phase are also intrinsic to the ED spectra. Thus our ED calculations confirm that all the universal features of the dynamical response observed in experiment are driven by the underlying spin-nematic order.


We close by turning our attention to the UUD-NSS transition. Because the NSS phase at zero field is also a spin nematic, the spectra shown in Fig.~\ref{fig2}D exhibit qualitatively the same low-energy characteristics as in Fig.~\ref{fig2}C, which are common to all pair condensates. Although a quantitatively accurate GSWT analysis is precluded in this case, our ED calculations demonstrate again that all the universal dynamical features are present in the one Goldstone and three single-magnon modes [Fig.~\ref{fig2}D]. Beyond this physics, we observe that the NSS phase displays further spectral features that break the direct relation to the single-magnon modes of the UUD phase, such as the resonance highlighted by the black box in Fig.~\ref{fig2}D2, which is also reproduced accurately in our ED calculations. The precise nature of these additional many-body phenomena remains as a target for further investigation.


To place our results in perspective, we have shown that spectroscopy provides an elegant means for probing directly the physical properties of pair condensates. For this purpose, our advantage over other pair-condensate platforms~\cite{stewart2008using,schirotzek2008,feld2011observation,hoinka2017goldstone, biss2022excitation} is that we can apply the full momentum- and energy-resolution of INS to measure not only the dispersion but also the decay characteristics of our quasiparticles. Less important in this regard is whether the condensate is fermionic or bosonic~\cite{anderson1995observation,bradley1995evidence, davis1995bose,zhang2021}, or aspects such as the spatial properties of the pairs, which can vary from highly delocalized and overlapping (BCS regime) to highly local (BEC regime)~\cite{randeria2014} without changing the universal dynamics. Nevertheless, the bosonic pair condensates known to date in quantum magnets tend to have rather local pairs~\cite{orlova2017,kohama2018,fogh2024b} and hence the band width of the Goldstone mode is very small. This combination of flat bands and low energy scales raises the possibility that the finite temperature of our experiments can cause thermal occupation of quasiparticle states and affect the field-induced quantum phase transitions. Indeed our experiments contain some evidence of a quantum critical regime~\cite{Sachdev2008} at intermediate fields through the FP-FQ and UUD-NSS transitions (Fig.~\ref{fig3}), and of thermal Goldstone-mode occupation in the FP phase in Fig.~\ref{fig5}B. However, thermal broadening is maximal at the lowest energies, in complete contrast to the characteristic pair-condensate broadening we find in the FQ and NSS phases (Figs.~\ref{fig2} and \ref{fig4}B), which is zero at the band bottom with a very {\bf Q}-specific fingerprint above this. Thus we can conclude that temperature effects on the universal dispersion and decay features we measure are negligible. 


In summary, we have used high-resolution INS experiments on the $S = 1$ quantum magnet \NiTL{} to achieve a systematic characterization of the field-induced evolution of the spin excitation spectrum through four distinct magnetic phases that include two spin nematics. Although the spin-nematic phase, the pair condensate of bosonic magnons, has historically evaded direct detection through a local order parameter, we obtain unequivocal evidence for its existence by understanding its dynamical properties. In the process we have uncovered the universal dynamics of the charge-neutral pair condensate, where a gapless and linearly dispersive Goldstone mode determines the stability of a gapped single-particle mode. The phenomenon by which this mode decays above a critical velocity, as verified in our experimental measurements, represents a condensed-matter analog of the Cherenkov radiation observed when a superluminal charged particle travels through a dielectric medium. In this way our work underscores the pivotal role of dynamical studies in advancing the understanding of novel states in quantum matter. 

\clearpage

\newpage

\newpage

\newpage
 
\newpage

\clearpage

\section*{References and Notes}
\bibliographystyle{Science_allauthors}
\renewcommand\refname{\vskip -1cm}
\bibliography{reference}
\clearpage

\section*{Acknowledgments}

\noindent
We thank C. D. Batista, S.-K. Jian, M. Mourigal, and I. Vekhter for valuable discussions, C. Schmitt for technical support during the INS experiments, and A. T. Savici for assistance in data reduction.

\paragraph*{Funding:}This work was supported by the U.S. Department of Energy, Office of Science, Basic Energy Sciences, under Award Number DE-SC0025426, DE-SC0020254, DE-SC0018660, and DE-SC0022311. H. Zhang acknowledges with gratitude the support of the U.S.~Department of Energy through the LANL/LDRD Program and the Center for Non-Linear Studies. The work performed at NHMFL was supported by Grant No. NSF-DMR-1157490 and the State of Florida. Portions of this research were conducted with high performance computational resources provided by Louisiana State University (http://www.hpc.lsu.edu). This research used resources at the High Flux Isotope Reactor and the Spallation Neutron Source, DOE Office of Science User Facilities operated by the Oak Ridge National Laboratory. 

\paragraph*{Author contributions:} The theoretical framework was conceived by S.-S.Z. The experimental program was designed by Q.H., H.Zhou and X.B. Q.H., W.Y. and H.Zhou grew the crystals and collected a.c. susceptibility data with assistance from E.S.C. Q.H., W.Y., H.Zhou, and X.B.~performed neutron scattering experiments with assistance from Y.H., D.M.P., and A.A.A. Q.H. and X.B. performed data analysis using packages developed by K.B.~and H.Zhang. S.-S.Z.~established the GSWT and performed these calculations. A.M.L. performed exact diagonalization calculations. Q.H., B.N., A.M.L., X.B., and S.-S.Z. developed the physical interpretation and wrote the manuscript with input from all the authors. 

\paragraph*{Competing interests:} The authors declare no competing interests.

\paragraph*{Data and materials availability:} All data reported in this paper are archived at http://www.xxx.yyy.zzz.\\

\section*{Supplementary Materials}

\noindent
Materials and Methods

\noindent
Supplementary Text

\noindent
Figs.~S1 to S19

\noindent
Table~S1

\noindent
References ({\it {50--57}})

\clearpage


       \renewcommand\refname{References}
       \renewcommand{\thesection}{\arabic{section}}
       \renewcommand{\thesubsection}{\thesection.\arabic{subsection}}

        \setcounter{equation}{0}
        \makeatletter 
        \def\tagform@#1{\maketag@@@{(S\ignorespaces#1\unskip\@@italiccorr)}}
        \makeatother
        \setcounter{figure}{0}
        \makeatletter
        \makeatletter \renewcommand{\fnum@figure}
        {\figurename~S\thefigure}
        \makeatother
        \setcounter{table}{0}
        \makeatletter
        \makeatletter \renewcommand{\fnum@table}
        {\tablename~S\thetable}
        \makeatother

\newcommand*\mycaption[2]{\caption[#1]{#1#2}}
\setcounter{secnumdepth}{-2}

\begin{center}
\section*{Supplementary Materials to accompany the manuscript \\ Universal dynamics of a pair condensate}

Qing Huang$^{1,2}$,\, Hao Zhang$^{3}$, Yiqing Hao$^{4}$,\\ Weiliang Yao$^{2}$, Daniel M. Pajerowski$^{4}$,
Adam A. Aczel$^{4}$,\\  Eun Sang Choi$^{5}$, Kipton Barros$^{3}$, Bruce Normand$^{6,7}$,\\
Haidong Zhou$^{2,}$\footnote{Email: hzhou10@utk.edu}, Andreas M. L\"auchli$^{6,7,}$\footnote{Email: andreas.laeuchli@psi.ch}, Xiaojian Bai$^{1,}$\footnote{Email: xbai@lsu.edu}, Shang-Shun Zhang$^{2}$\footnote{Email: szhang57@utk.edu}\\
\vspace{0.5cm}
\normalsize{$^1$Department of Physics and Astronomy, Louisiana State University, Baton Rouge, LA 70803, USA} 
\\
\normalsize{$^2$Department of Physics and Astronomy, University of Tennessee, Knoxville, TN 37996, USA}
\\
\normalsize{$^3$Theoretical Division and CNLS, Los Alamos National Laboratory, Los Alamos, NM 87545, USA}
\\
\normalsize{$^4$Neutron Scattering Division, Oak Ridge National Laboratory, Oak Ridge, TN 37831, USA}
\\
\normalsize{$^5$National High Magnetic Field Laboratory, Florida State University, Tallahassee, Florida 32310-3706, USA}
\\
\normalsize{$^6$PSI Center for Scientific Computing, Theory and Data, CH-5232 Villigen-PSI, Switzerland}
\\
\normalsize{$^7$Institute of Physics, Ecole Polytechnique F\'ed\'erale de Lausanne (EPFL), CH-1015 Lausanne, Switzerland}
 
\end{center}

\tableofcontents
\vspace{0.5cm}

\section{S1   Materials and Methods}
\subsection{S1.1   Experimental Details}

Single-crystal samples of Na$_2$BaNi(PO$_4$)$_2$ were synthesized using an NaCl flux, following the procedure detailed in Ref.~\cite{li2021a}. Approximately 5~g of plate-like crystals were coaligned in the ($HK0$) scattering plane for our inelastic neutron scattering (INS) measurements. A picture of the crystal mount is shown in Fig.~S\ref{SI_crystal}A.  

For characterization purposes, a.c. susceptibility ($\chi'$) measurements were performed using SCM1 at the National High Magnetic Field Laboratory in Tallahassee, which is equipped with a top-loading dilution refrigerator capable of reaching a base temperature of 20 mK. The measurement was conducted using an a.c.~excitation field with an RMS amplitude of 0.6 Oe and a fixed frequency of 220 Hz. The data in Fig.~2B1 were collected at $T = 50$~mK with the magnetic field applied along the $c$ axis of the crystal. A complete a.c.~susceptibility dataset from $T = 0.02$ to $0.94$~K is presented in Ref.~\cite{li2021a}.

INS experiments were performed on the CNCS spectrometer at the Spallation Neutron Source (SNS) at Oak Ridge National Laboratory (ORNL), USA~\cite{ehlers2011}. The sample was mounted on a dilution refrigerator insert, placed inside an 8~T vertical-field self-shielded superconducting magnet, and cooled to a base temperature of $T \simeq 50$~mK. Two separate experiments were performed using the same crystal mount and sample environment. The first dataset was collected by rotating the sample around the vertical axis in $1^\circ$ steps, covering a total range of $180^\circ$ to map a large area of reciprocal space at the applied magnetic fields $B = 0.0$, 0.2, 1.0, 1.5, 1.8, and 4.0~T. To track spectral changes in detail, a second experiment was performed focusing on a 30$^\circ$ range centered on the K point of the Brillouin zone (BZ), using small field increments to traverse the transitions from the FP to the FQ phase ($B = 2.0$ to 1.7~T) and from the UUD to the NSS phase ($B = 0.2$ to 0.0~T). 

An additional dataset was collected at high field ($B = 5.0$~T) over the same angular range for the same counting time, and this alone was used as the instrumental background we subtracted (i.e.~no further potential background contributions were estimated) to reveal the presence of the Goldstone mode at very low energy in Fig.~3 of the main text. An incident neutron energy of $E_\text{i} = 1$~meV was used to achieve the best possible energy resolution of the low-energy excitations, specifically an FWHM elastic resolution of approximately 0.025~meV: this capability was the key that enabled us to measure both the dispersion and the decay of the single magnon to high accuracy over the whole BZ. Additional data were collected with an incident energy of $E_\text{i} = 12$~meV to characterize the field-dependence of the magnetic Bragg-peak intensities and the results of these measurements are presented in Figs.~S\ref{SI_Bragg} and S\ref{SI_OP}.

Unpolarized neutron scattering probes the dynamical structure factors $S^{\alpha\beta}(\bQ,E)$ for spin components $\alpha, \beta = x, y, z$ perpendicular to the momentum transfer $\bQ$, 
\begin{align}
    \label{eq:ins}
    \mathcal{S}(\bQ,E) &= \sum_{\alpha\beta} \left(\delta^{\alpha\beta} - \frac{Q^\alpha Q^\beta}{Q^2} \right) S^{\alpha\beta}(\bQ,E) \nonumber \\
    &= \left(1 - \frac{Q^x Q^x}{Q^2} \right) S^{xx}(\bQ,E) + \left(1 - \frac{Q^y Q^y}{Q^2} \right) S^{yy}(\bQ,E) + S^{zz}(\bQ,E) \nonumber \\
    &= \left(2 - \frac{Q^x Q^x+Q^y Q^y}{Q^2} \right) S^{xx}(\bQ,E) + S^{zz}(\bQ,E) \nonumber \\
    &= S^{xx}(\bQ,E) + S^{zz}(\bQ,E),
\end{align}
where in the second to fourth lines we have applied the conditions that $\bQ$ is in the horizontal scattering plane and that the vertical field does not lower the symmetry of the XXZ Hamiltonian [Eq.~(1) of the main text], whence $S^{xx}(\bQ,E) = S^{yy}(\bQ,E)$. All of our INS data were reduced and analyzed using MANTID~\cite{arnold2014} on the SNS analysis cluster at ORNL. The symmetry operations of the $\overline{3}$ point group were applied to the data to increase the statistics. Figure~S\ref{SI_crystal}B defines the high-symmetry points of the BZ labeled in the main-text figures, $\Gamma$, Y, K, M, X, and X$'$. GSWT fits to the INS spectra in the FP phase were performed using the \textsc{Sunny.jl} package \cite{zhang2021b,dahlbom2022f,Sunny2025}.

\subsection{S1.2   Generalized Spin-Wave Theory}

To describe the spin dynamics in the FP and UUD phases, we applied a standard $\mathrm{SU}$(3) spin-wave theory appropriate for $S = 1$ spins~\cite{muniz2014generalized,do2021decay,bai2023instabilities} to obtain an accurate account of the experimental spectra (Fig.~2C of the main text). However, a direct application of this approach to the FQ and NSS phases is not possible because neither phase is a stable classical ground state of the spin Hamiltonian. Their emergence is a consequence of quantum fluctuation effects that confer dynamical properties on the TMBS, as represented in Fig.~4A1 of the main text. These are captured by a renormalized Hamiltonian, where Eq.~(1) of the main text is augmented by introducing a hopping term ${\cal H}_{\rm TMBS}  = \sum_{ij} \tfrac{1}{2} \delta t_{ij} (S_i^+)^2 (S_j^-)^2$; for the large $D/J$ ratio relevant in \NiTL, $\delta t_{ij}$ is dominated by the second-order process, making it negative and ${\cal {O}} (J^2/D)$ in magnitude. A full derivation is presented in Sec.~S2.2. The TMBS described by the renormalized Hamiltonian is then a dispersive mode with a minimum at the $\Gamma$ point, where it condenses as the field is reduced to the FP-FQ transition. This condensation induces a quadrupole moment in the ground state without further breaking of translational symmetry. Because this quadrupolar phase is the classical ground state of the renormalized Hamiltonian, we can again apply a conventional $\mathrm{SU}$(3) spin-wave theory to perform a systematic study of its quantum fluctuations, and in this way we computed the dispersion and damping shown in Fig.~2C2 of the main text.

\subsection{S1.3   Exact Diagonalization}

ED provides the exact eigenvalues and eigenstates of the quantum Hamiltonian for any system, regularizing the size of the matrix to be diagonalized simply by limiting the number of lattice sites. A $S = 1$ system has $3^N$ states on an $N$-site cluster, and we limit the matrix sizes by exploiting the global U(1) symmetry of the Hamiltonian in Eq.~(1) of the main text to work in sectors of fixed $S_z$, which we then convert to discrete effective system magnetizations. We also use lattice translation symmetries in order to work at fixed total momentum. The accessible cluster size then increases with $S_z$, allowing us to compute the low-lying energy spectra on cluster sizes ranging from $N = 12$ to $24$ at zero field and from $12$ to $36$ at high fields, a size range being required to analyze the convergence of the results. We use Krylov techniques to compute the quantities $S^{xx}(\bQ,E)$ and $S^{zz}(\bQ,E)$ of Eq.~\eqref{eq:ins}, and in Figs.~2 and 5 of the main text we present the results over a continuous range of $E$ after convolution with the instrumental resolution, but for a restricted set of symmetric wave vectors $\bQ$ allowed by the size and shape of the cluster (shown in Fig.~S\ref{SI_crystal}B). Further details of our ED methods and spectra are presented in Sec.~S2.3.

\clearpage

\section{S2   Supplementary Text and Figures}

\subsection{S2.1   Supplementary Neutron-Scattering Data}

\begin{figure}[h!] 
 	\begin{center}
 		\includegraphics[width=0.6\textwidth]{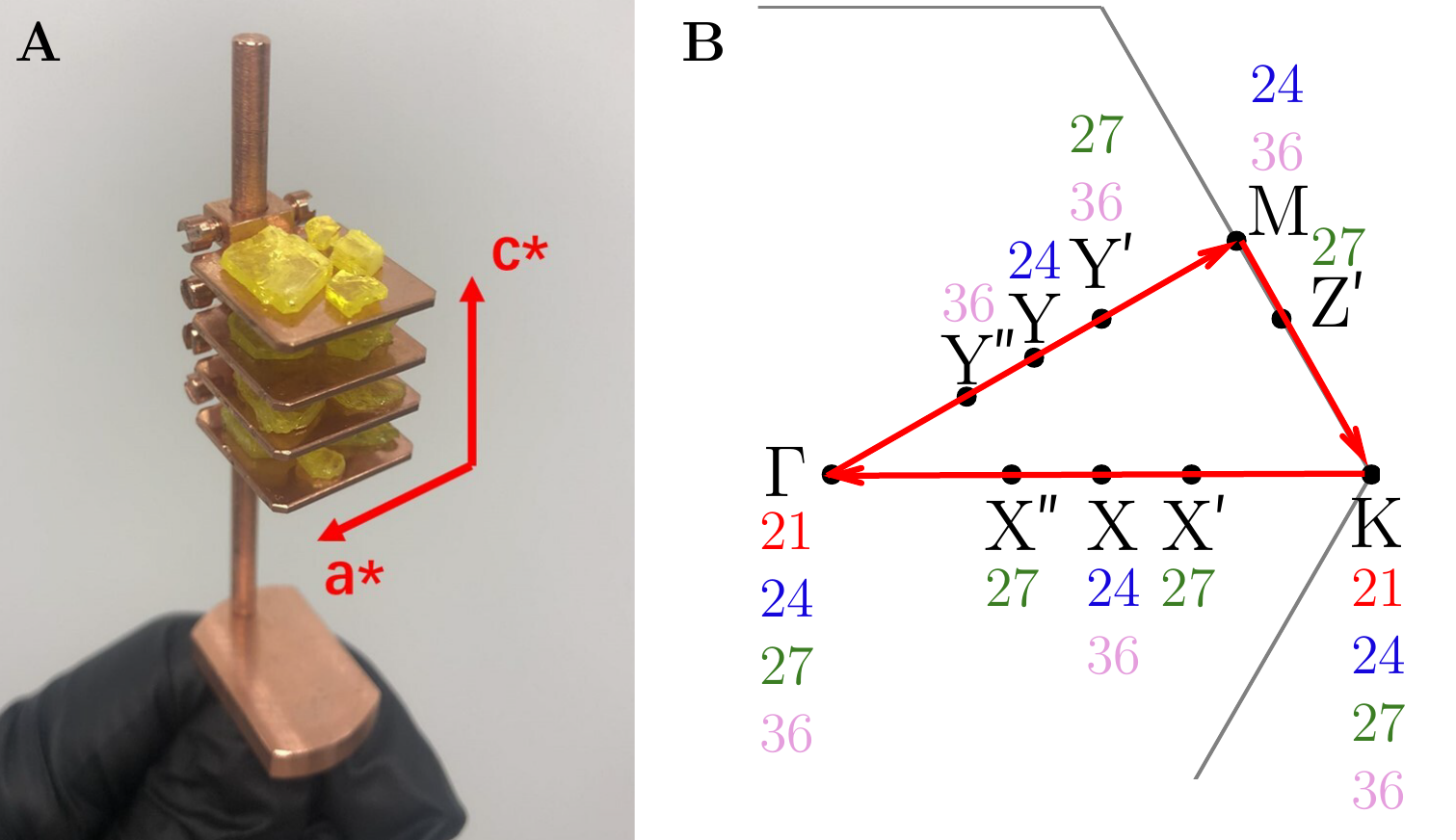}
	\end{center}
\caption{\label{SI_crystal} ({\bf A}) Crystal mount, showing aligned \NiTL~crystallites totaling approximately 5~g in mass. ({\bf B}) Representation of the high-symmetry path in the Brillouin zone along which the dispersion relations measured by INS are shown, with specific high-symmetry points marked. The numbers represent cluster sizes in our ED calculations on which specific {\bf Q} points are accessible.}
\end{figure}

\begin{figure}[h!] 
 	\begin{center}
 		\includegraphics[width=0.8\textwidth]{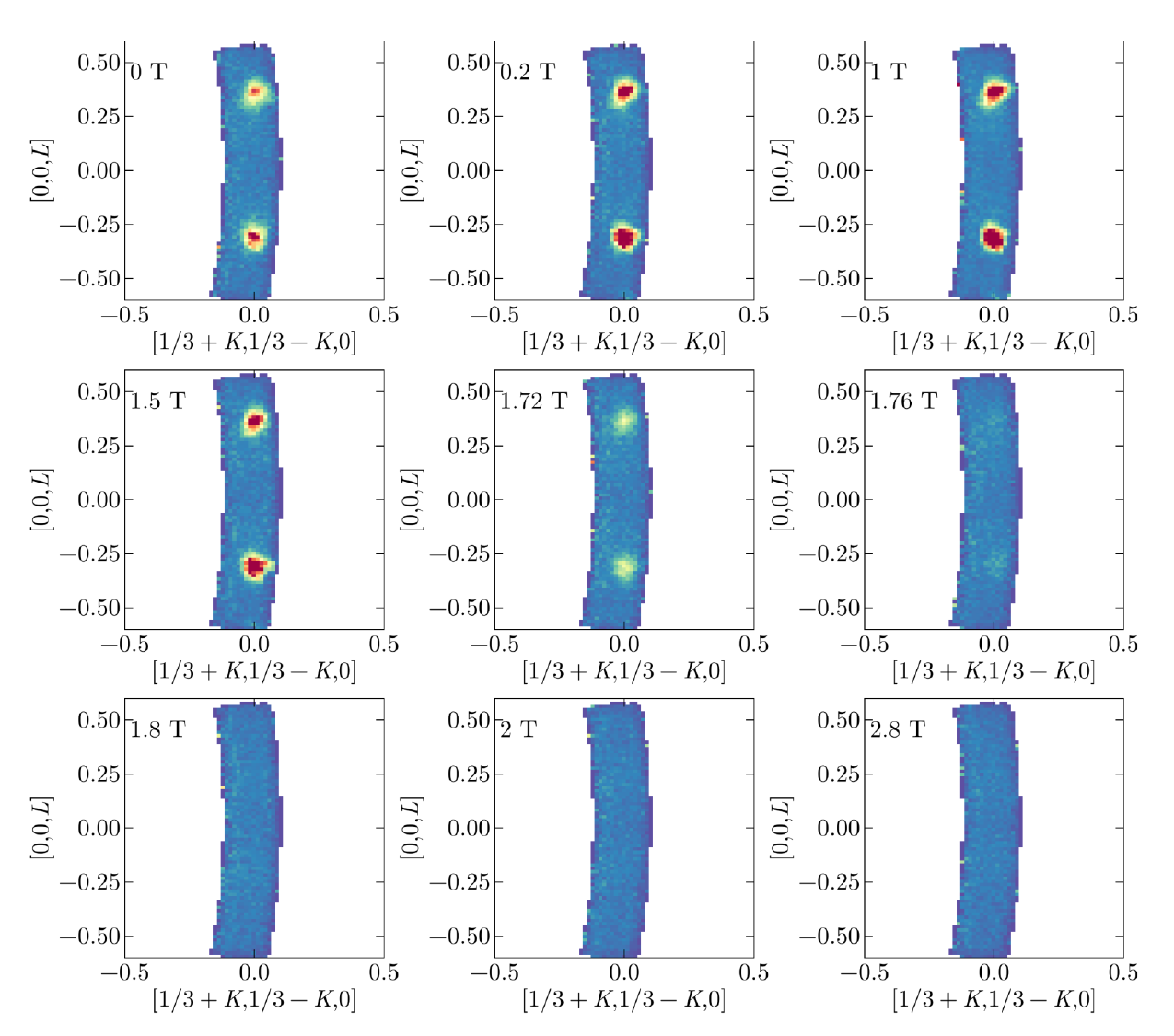}
	\end{center}
\caption{\label{SI_Bragg} Magnetic Bragg peaks shown as constant-energy scans in the ($1/3+K$ $1/3-K$ $L$) plane at the elastic line, taken from our CNCS data collected over a wide range of applied magnetic fields with incident energy $E_\text{i} = 12$ meV.}
\end{figure}

\begin{figure}[h!] 
 	\begin{center}
 		\includegraphics[width=0.6\textwidth]{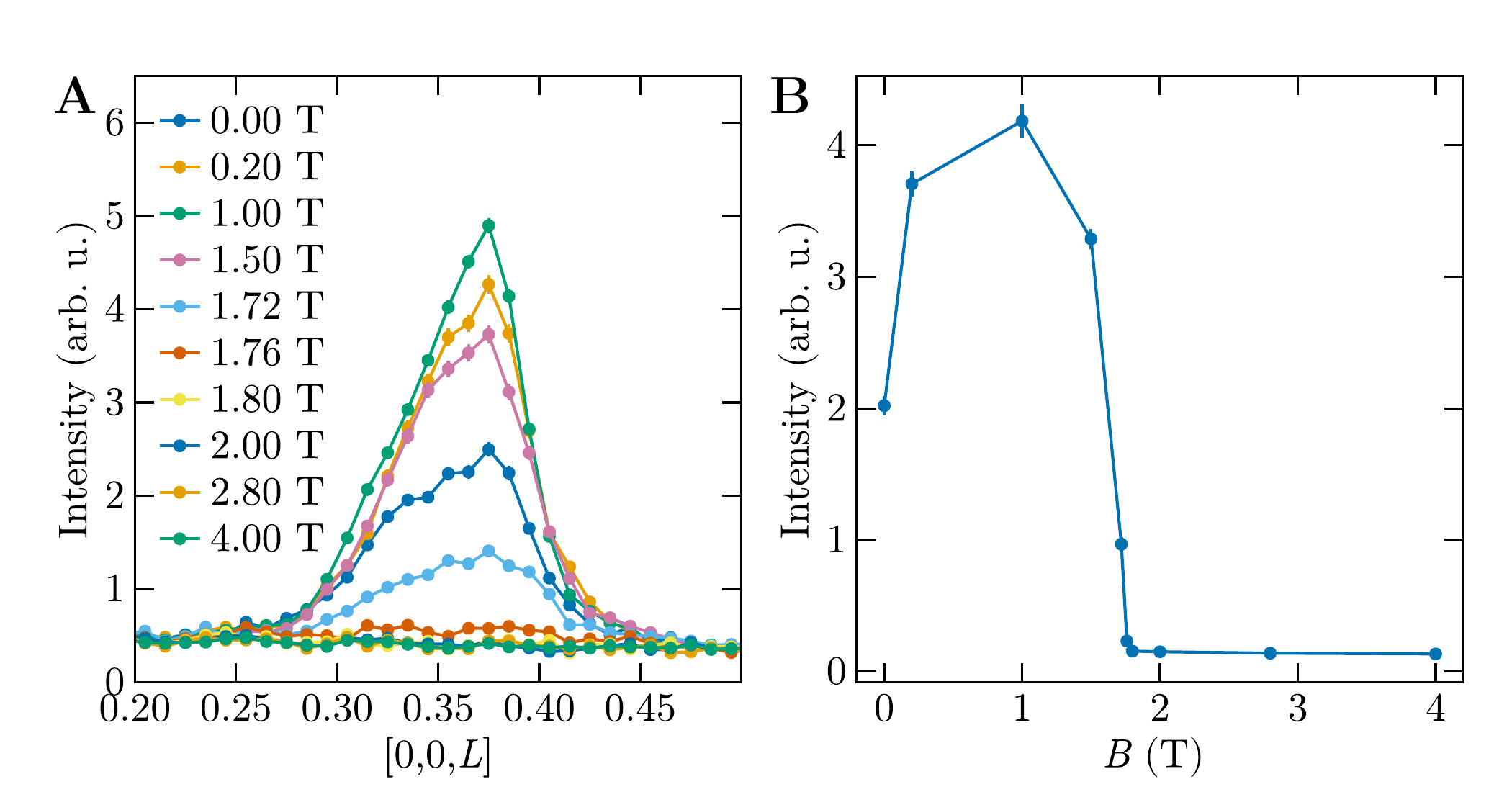}
	\end{center}
\caption{\label{SI_OP} ({\bf A}) Profile of the magnetic Bragg peak along the [1/3,1/3,$L$] direction taken from the data of Fig.~S\ref{SI_Bragg}. ({\bf B}) Field-dependence of the magnetic order parameter extracted from the data of panel (A).}
\end{figure}

\begin{figure}[h!] 
 	\begin{center}
 		\includegraphics[width=0.6\textwidth]{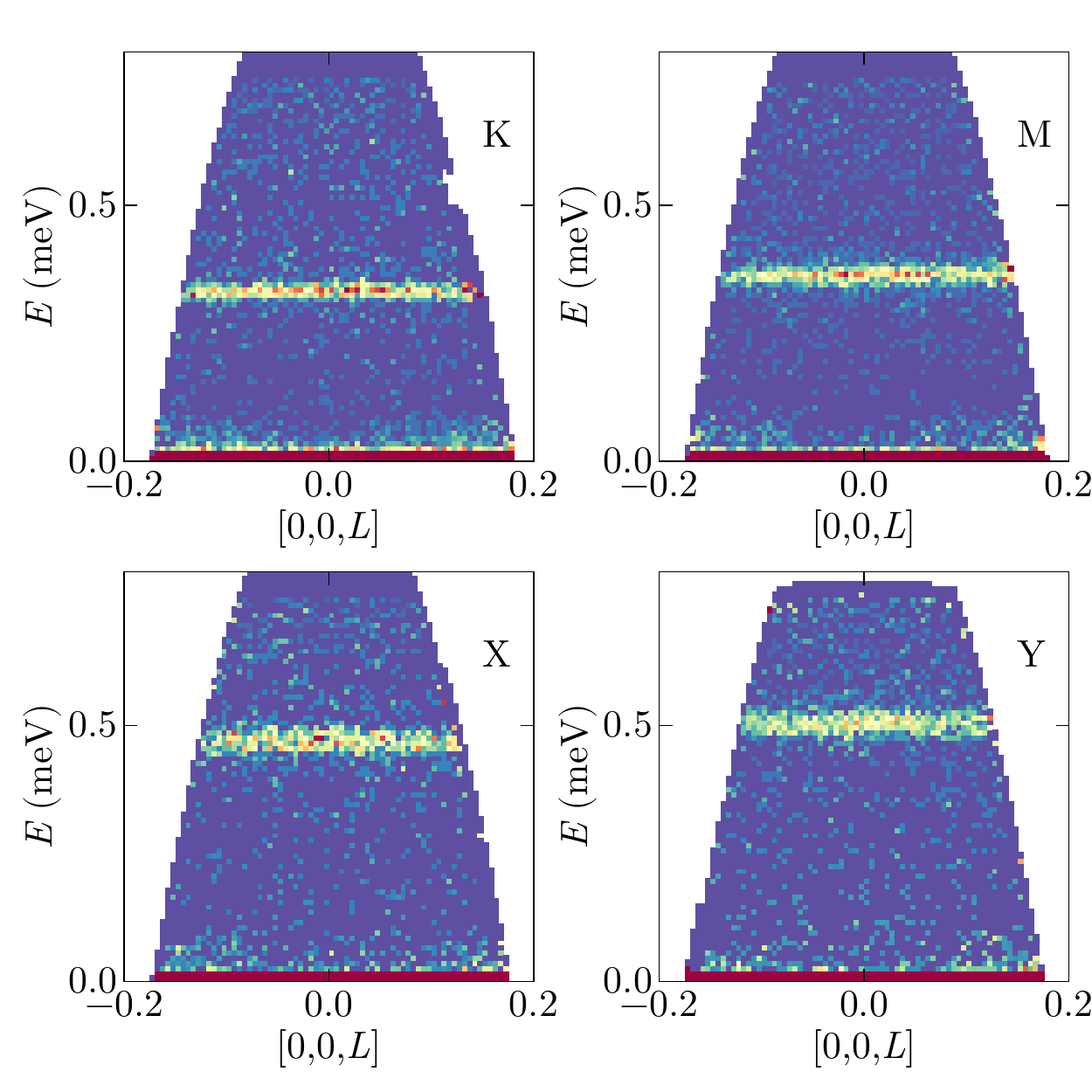}
	\end{center}
\caption{\label{SI_out_of_plane} Illustration of the entirely flat single-magnon dispersion in the $L$ direction, measured at K, M, X and Y under a magnetic field of 4~T.}
\end{figure}

For a graphical illustration of the foundations of our INS experiments, in Fig.~S\ref{SI_crystal}A we show a photograph of the sample prior to loading in the cryomagnet and mounting in the CNCS spectrometer. Figure S\ref{SI_crystal}B shows the Brillouin zone of the triangular lattice, highlighting the high-symmetry path in reciprocal space over which we prepared the {\bf Q}-dependent spectra shown in Figs.~2 and 3 of the main text. To characterize the ground-state phase diagram as a function of the applied field, in Fig.~S\ref{SI_Bragg} we show the magnetic Bragg peaks associated with antiferromagnetic order, which breaks the translational symmetry of the triangular lattice, measured over the full field range. In Fig.~S\ref{SI_OP} we compare the peak intensities, which highlight the absence of spontaneous long-ranged magnetic order in the FQ and FP phases, i.e.~above the UUD-FQ transition at $B = 1.76$~T, and the reduction of the dipolar order parameter as the system passes from the UUD to the NSS phase at very low fields. Although this bulk magnetic order must be established by an interplane interaction, Fig.~S\ref{SI_out_of_plane} demonstrates that the single-magnon excitations have absolutely no discernible dispersion along the out-of-plane direction, even on the scale of our very high measurement resolution. Thus any interplane interaction must be extremely small, justifying that the minimal model should contain only intra-layer coupling terms [Eq.~(1) of the main text].

\subsection{S2.2   Formulation of SU(3) generalized spin-wave theory}

\begin{figure}[h!] 
 	\begin{center}
 		\includegraphics[width=0.9\textwidth]{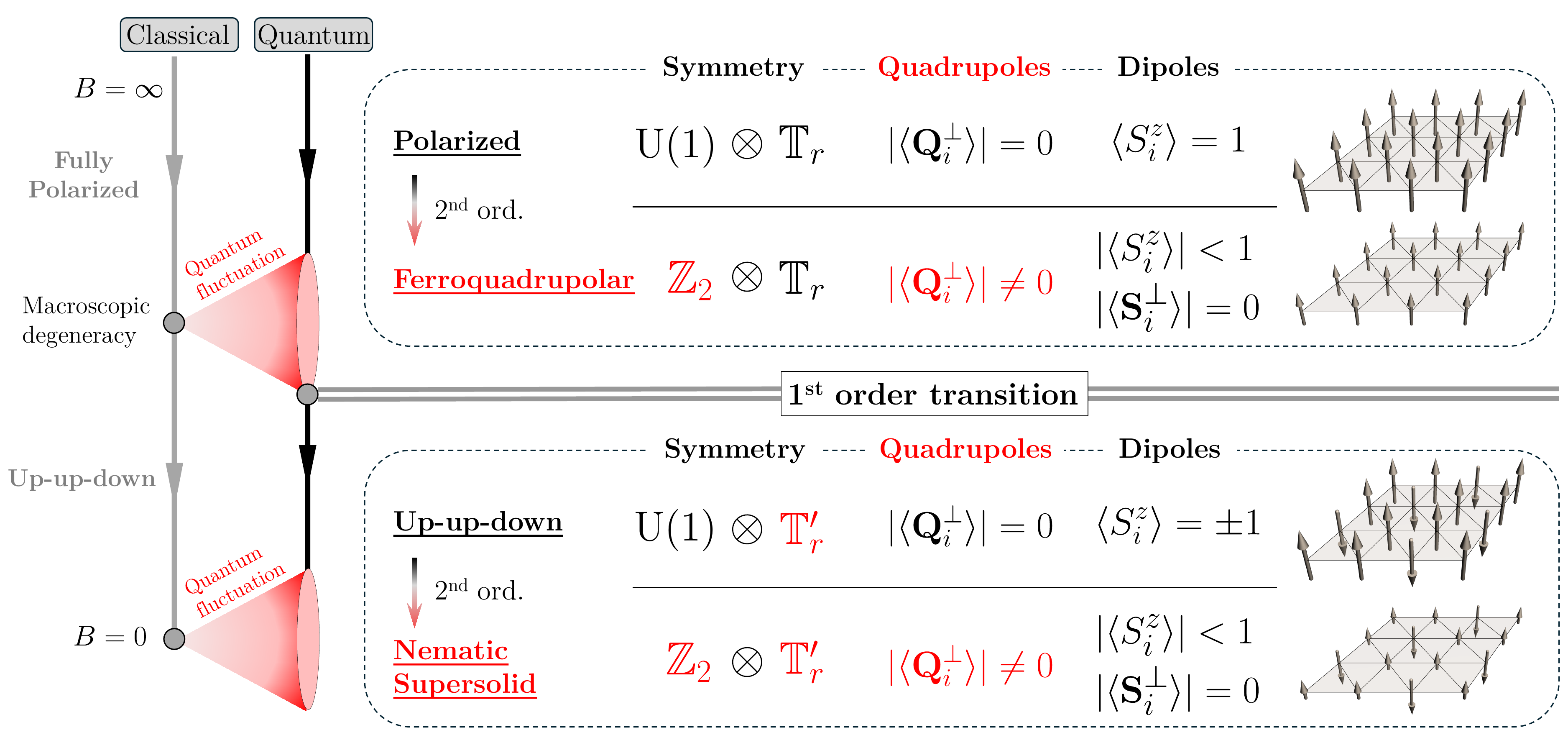}
	\end{center}
\caption{\label{SI_phase_diagram} Classical and quantum phase diagrams of the model Hamiltonian for \NiTL~given in Eq.~(1) of the main text. In each of the four phases of the quantum system, we indicate the relevant symmetry groups and the nature of the quadrupolar and dipolar order parameters [the former defined in Eq.~\eqref{eq:order_parameter}]. ${\mathbb T}_r$ denotes the primitive translation group of the triangular lattice and ${\mathbb T}_r^\prime$ its reduced form in the UUD and NSS phases. ${\mathrm U}(1)$ is the spin-rotation group about the field axis and ${\mathbb Z}_2$ the reduced group corresponding to a $\pi$ rotation.}
\end{figure}

The phase diagram of the model given in Eq.~\eqref{eq:model} with classical spins has two collinear magnetically ordered ground states, the fully polarized (FP) state and the up-up-down (UUD) state, as shown n the left side of Fig.~S\ref{SI_phase_diagram}. Both phases preserve the ${\mathrm U}(1)$ spin rotation symmetry about the field direction and differ only in their translational symmetry and their excitations are well described by a standard ${\mathrm{SU}}(3)$ GSWT. However, many-body quantum fluctuations cause two additional phases to appear, one near the FP-UUD transition of the classical system and one close to zero field. In both of these phases, the ${\mathrm U}(1)$ symmetry is spontaneously down to a discrete ${\mathbb Z}_2$ symmetry, which corresponds to a $\pi$ spin rotation about the field axis. This symmetry-breaking is manifest as a ground state with no transverse dipolar order, $\langle \mathbf{S}_i^\bot \rangle = (\langle S_i^x\rangle, \langle S_i^y\rangle) = 0$, but a finite quadrupole moment,
\begin{equation}
\label{eq:order_parameter}
\mathbf{Q}_i^\bot = \bigl(\mathrm{Re}[(S_i^+)^2], \, \mathrm{Im}[(S_i^+)^2]\bigr) \;\; \neq \;\; 0.
\end{equation}
Because these two quadrupolar phases, the ferroquadrupolar (FQ) and nematic supersolid (NSS) phases, have no classical counterparts, the standard SU(3) GSWT is not directly applicable to either. To describe these states, we develop a renormalized GSWT applicable to situations with quadrupolar order, meaning that this type of order is stabilized in the renormalized semiclassical limit. In this section, we first provide a summary of the standard SU(3) GSWT framework and then introduce the renormalization approach that allows the description of the FQ and NSS phases within the same framework.

\subsubsection{Standard SU(3) GSWT -- general formulation}

The ${\mathrm{SU}}(3)$ GSWT is constructed by associating the three low-energy crystal-field levels (Fig.~2A of the main text), which we denote as \( \rvert \sigma \rangle \) with three boson flavors, \( \sigma = +,0,- \), according to 
\begin{eqnarray}\label{eq:basis}
    \rvert\sigma\rangle = b_{\sigma}^\dagger \rvert 0 \rangle,
\end{eqnarray}
where \( \rvert 0 \rangle \) refers to the vacuum of Schwinger bosons~\cite{muniz2014generalized,do2021decay,bai2023instabilities}. To guarantee the local dimensionality of a spin-$1$ operator, these bosons satisfy the local constraint 
\begin{eqnarray}\label{eq:constraint}
    \sum_{\sigma=\pm,0} b_{\sigma}^\dagger b_{\sigma} = M,
\end{eqnarray}
where \( M = 1 \) is associated with the fundamental irreducible representation of the ${\mathrm{SU}}(3)$ group. Computing the matrix elements of the spin and quadrupole operators in the basis of Eq.~\eqref{eq:basis} yields the Schwinger-boson representation of the spin operators,
\begin{align}
\hat{S}_{i}^{+} = \sqrt{2} \left( b_{i,+}^{\dagger} b_{i,0} + b_{i,0}^{\dagger} b_{i,-} \right), \quad
\hat{S}_{i}^{-} = \left( \hat{S}_{i}^{+} \right)^{\dagger}, \quad
\hat{S}_{i}^{z} = b_{i,+}^{\dagger} b_{i,+} - b_{i,-}^{\dagger} b_{i,-},
\end{align}
while the quadrupole operator, which appears in the single-ion term of the starting Hamiltonian, is  
\begin{align}
(S_{i}^{z})^{2} = b_{i,+}^{\dagger} b_{i,+} + b_{i,-}^{\dagger} b_{i,-}.
\end{align}
In this way the spin Hamiltonian given in Eq.~(1) of the main text is mapped to a Schwinger-boson Hamiltonian subject to the local constraint of Eq.~\eqref{eq:constraint}.

SWT is a perturbation theory controlled by the small parameter \( 1/M \), where \( M \to \infty \) is the classical limit in which the spin operators are replaced by vectors. The classical ground state is a product state, \( \otimes_i \rvert \phi_i \rangle \), where \( \rvert \phi_i \rangle \) denotes the coherent state of a boson at lattice site \( i \), which can be expanded in the form \( \rvert\phi_i\rangle = \sum_\sigma \phi_{i,\sigma} \rvert \sigma \rangle \) to include the local constraint [Eq.~\eqref{eq:constraint}]. The energy of this product state is given by replacing the operators \( b_{i\sigma}, b^\dagger_{i\sigma} \) with numbers \( \phi_{i,\sigma}, \phi^*_{i,\sigma} \) in the Schwinger-boson Hamiltonian, and minimizing this classical energy with respect to \( \phi_{i,\sigma} \) yields the classical ground state. 

To study quantum fluctuation effects, it is convenient to transform from the spin reference frame to a local Schwinger-boson frame, defined by a unitary transformation
\begin{eqnarray}
    (b_{i,+}, b_{i,0}, b_{i,-})^T  = U_i (\beta_{i,+}, \beta_{i,0}, \beta_{i,-})^T, 
\end{eqnarray}
where the first column of \( U_i \) is given by the normalized vector \( (\phi_{i +}, \phi_{i 0}, \phi_{i -})^T \) and the other two columns are orthogonal to the first one. The defining property of this local reference frame is that magnetic order is described by condensation in the single-particle state \( \beta_{i,+}^\dagger \rvert 0 \rangle \), while the other two boson flavors, \( \beta_{i,0} \) and \( \beta_{i,-} \), describe quantum fluctuations. Because this transformation is unitary, the local constraint remains form-invariant, i.e.~\( \sum_{\sigma = \pm,0} \beta_{\sigma}^\dagger \beta_{\sigma} = M \). Expanding the condensate fraction as a perturbation series in \( 1/M \) yields 
\begin{equation}
\beta_{i,+} = \beta_{i,+}^\dagger = \sqrt{M - \beta^\dagger_{i,0} \beta_{i,0} - \beta_{i,-}^\dagger \beta_{i,-}} \simeq \sqrt{M} - \frac{1}{2\sqrt{M}} \beta_{i,0}^\dagger \beta_{i,0} - \frac{1}{2\sqrt{M}} \beta_{i,-}^\dagger \beta_{i,-} + \dots.
\end{equation}
Expanding the original spin Hamiltonian in the same way leads to the form 
\begin{equation}\label{eq:Hsw}
{\cal H} = M^2 {\cal H}_0 + M {\cal H}_2 + M^{1/2}{\cal H}_3 +  M^0 {\cal H}_4 + \dots,
\end{equation}
where the first term gives the classical interaction energy and all terms \( {\cal H}_n  \) with $n > 1$ contain \( n \) boson operators. These have the straightforward physical interpretation that ${\cal H}_2$ is the GSWT Hamiltonian that describes the kinematic spectrum of non-interacting quasiparticles and higher terms describe the interactions between quasiparticles. With ${\cal O}(1/M)$ accuracy, ${\cal H}_3$ is responsible for quasiparticle decay and ${\cal H}_4$ for Hartree-Fock energy renormalization. At the end of the calculation one takes $M \rightarrow 1$ to obtain the physical situation.

As noted above, this formulation describes the FP and UUD phases, whereas the FQ and NSS phases are stabilized by quantum fluctuations of the type illustrated schematically in Fig.~\ref{fig4}{\bf A2} of the main text. To formulate a semiclassical framework taking such physical processes into account, we proceed in two steps. First, as summarized in the main text, the physics of higher-order quantum fluctuations is that the TMBS acquires a dispersion and a field-induced condensation at its band minimum generates the quadrupole moments of the FQ and NSS phases. Second, we incorporate this phenomenon into a renormalized Hamiltonian, which enables the description of the FQ and NSS phases near the phase transition points with the $1/M$ expansion.

\subsubsection{Renormalized Hamiltonian for FP and FQ phases}

The FP phase corresponds to a condensate of \( b_{i,+} \) bosons at each lattice site \( i \), whence the local reference frame coincides with the original one. Following the \( 1/M \) expansion scheme, the quadratic GSWT Hamiltonian is
\begin{align}\label{eq:h2}
{\cal H}_{2} = J \sum_{\langle ij\rangle} \left( b_{j,0}^{\dagger} b_{i,0} + {\rm h.c.} \right) + \left( h + D - 6 J {{\Delta}} \right) \sum_{i} n_{i,0} + \left( 2 h - 12 J {{\Delta}} \right) \sum_{i} n_{i,-},
\end{align}
where \( n_{i,\sigma} = b_{i,\sigma}^\dagger b_{i,\sigma} \). After Fourier transformation one obtains the single-magnon dispersion
\begin{align}\label{eq:smd}
\varepsilon_{0,{\bm k}} = 2 J \gamma_{\bm k} + h + D - 6 J {\Delta},
\end{align}
where
\begin{equation}
    \gamma_{\bm k} = \sum_{{\bm \delta} > 0} \cos({\bm k} \cdot {\bm \delta}),
\end{equation}
with \( {\bm \delta} = (a,0), (\pm {a/2},\sqrt{3}a/2) \) the nearest-neighbor bonds on the triangular lattice and \( a \) the lattice constant. This dispersion has a global minimum at the K point of the primitive Brillouin zone, \( \varepsilon_{0,K} = h + D - 9J {{\Delta}} \). The expression of Eq.~\eqref{eq:smd} is an exact result, which we used to compute the sharp single-magnon band in Fig.~2C1 of the main text and to extract the best-fit model parameters for \NiTL (Sec.~S2.5). 

\begin{figure}[h!] 
 	\begin{center}
 		\includegraphics[width=0.7\textwidth]{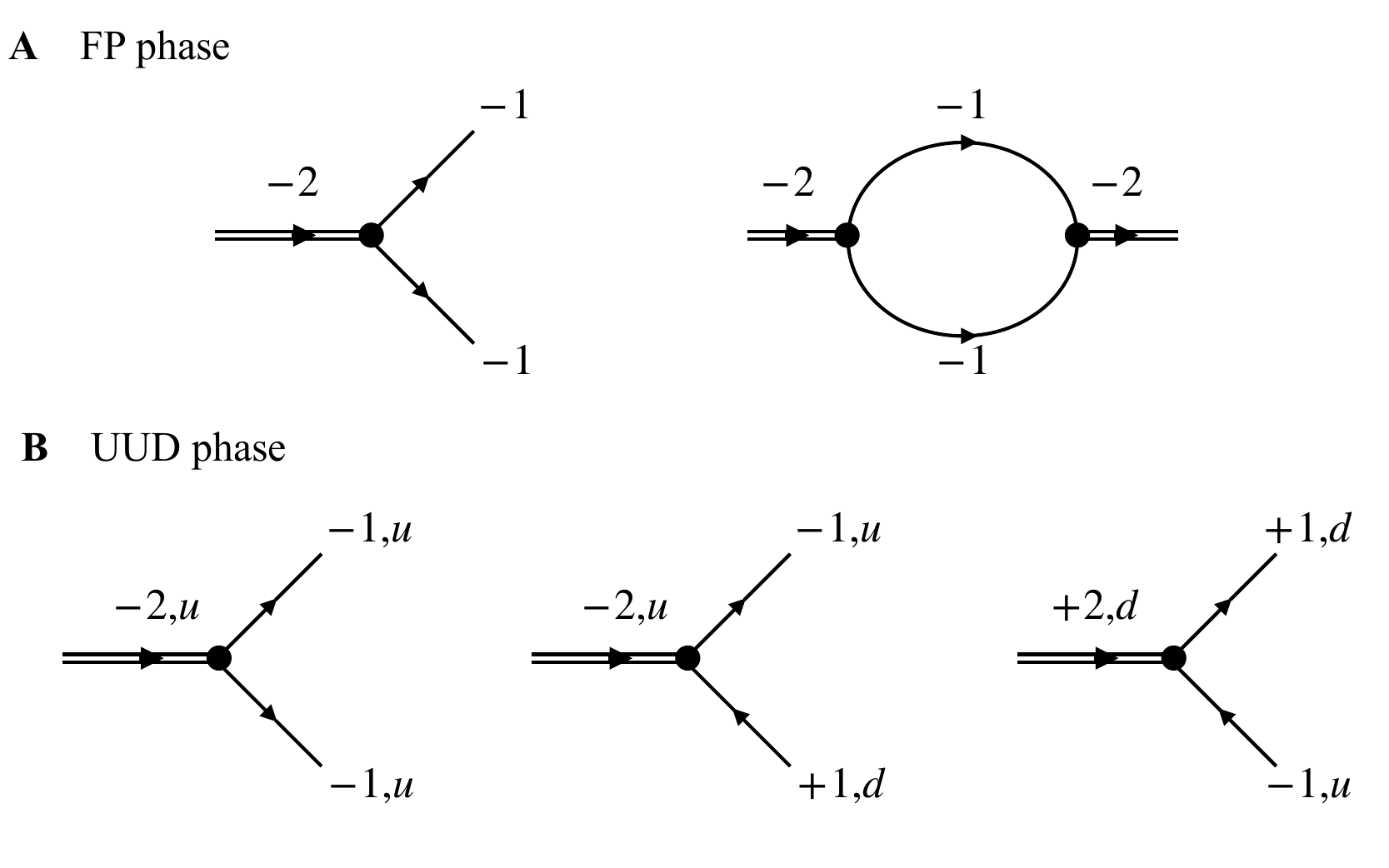}
	\end{center}
\caption{\label{SI_vertex} ({\bf A}) Cubic vertex in the FP phase (left) and one-loop self-energy of the two-magnon bound state (right), where the latter corresponds to the physical process depicted in Fig.~4A2 of the main text. Single lines denote a single-magnon propagator and double lines a TMBS propagator. Numbers label the spin quantum number carried by each propagator along the quantization axis defined by the external field direction. ({\bf B}) Cubic vertex in the UUD phase, where $u$ and $d$ label the up and down sublattices on which the local spin-flips appear. The one-loop self-energy of the two-magnon bound state is the same as in the FP phase if one adopts a Nambu representation [where $\psi_{\bf k}$ describing the single-magnon excitations is defined below Eq.~\eqref{eq:si_cubic_uud}].}
\end{figure}

By the definition of the three Schwinger-boson states as crystal-field levels, $b_{i,-}^{\dagger}$ is the creation operator of a TMBS, and from the last term of Eq.~\eqref{eq:h2} one observes that the TMBS dispersion,
\begin{eqnarray}
    \varepsilon_{-,{\bm k}} = 2 h - 12 J {{\Delta}},
\end{eqnarray}
is entirely flat. In the FP phase, both the single magnon and the TMBS are gapped, and when \( D > 3 J \Delta \), which is relevant for Na$_2$BaNi(PO$_4$)$_2$, the TMBS lies lower, becoming gapless at an applied field of \( h_c = 6 J \Delta \) while the single magnon remains gapped. The ground-state instability of the FQ phase is then a quadrupolar ordering driven by field-induced TMBS condensation. To determine the ordering wavevector of the TMBS condensate, we include the next-order quantum processes represented graphically Fig.~\ref{fig4}A2 of the main text, which lead to a dispersive TMBS spectrum described by the renormalized Hamiltonian introduced in Sec.~S1.2. This effect has its origin in cubic interaction term, \( {\cal H}_3 \), which takes the form
\begin{eqnarray}
{\cal H}_{3} & = & J \sum_{\langle ij \rangle} b_{j,-}^{\dagger} b_{j,0} b_{i,0} + b_{i,-}^{\dagger} b_{i,0} b_{j,0} + {\rm h.c.} \nonumber \\
& = & \frac{1}{2 \sqrt{N}} \sum_{kq} V(k,q) b_{q,-}^{\dagger} b_{k,0} b_{q-k,0} + {\rm h.c.},
\end{eqnarray}
where $b_{{\bm k},\sigma} = (1/\sqrt{N}) \sum_{i} b_{i,\sigma} e^{-i{\bm k} \cdot {\bm r}_i}$ with $\sigma = 0,-$, ${\bm r}_i$ the lattice vector of site $i$, and \( V(k,q) = 2 J (\gamma_{\bm k} + \gamma_{{\bm q}-{\bm k}}) \) is the cubic vertex function (Fig.~S\ref{SI_vertex}A). Including the energy correction given by the one-loop self-energy diagram of Fig.~S\ref{SI_vertex}A, the TMBS dispersion becomes 
\begin{equation}
\tilde{\varepsilon}_{-,{\bm q}} = \varepsilon_{-,{\bm k}} - J \int \frac{d^2{\bm k}}{(2\pi)^2} \frac{ (\gamma_{\bm k} + \gamma_{{\bm q}-{\bm k}})^2 }{(\gamma_{\bm k} + \gamma_{{\bm q}-{\bm k}}) + D/J}.\label{eq:deltaE}.
\end{equation}
The correction is largest and negative at the \( \Gamma \) point (${\bm q} = {\bm 0}$), giving an estimate of the critical field at the lowest nontrivial order as
\begin{eqnarray}\label{eq:hc-gswt}
    h_{c} = 6 J {{\Delta}} + J \int \frac{d^2{\bm k}}{(2\pi)^2} \frac{ 2 \gamma_{\bm k}^2 }{2 \gamma_{\bm k} + D/J}.
\end{eqnarray}

The renormalized TMBS spectrum is that of the real-space tight-binding Hamiltonian
\begin{eqnarray}\label{eq:hopping_fq}
    \tilde{H}_{2,{\rm TMBS}} = \sum_{ij} (\delta t_{ij} \beta_{j,-}^\dagger\beta_{i,-} + h.c.) + (2h - \delta \mu) \sum_{i} \beta_{i,-}^\dagger \beta_{i,-},
\end{eqnarray}
where
\begin{align}\label{eq:tij}
\delta t_{ij} = \int{d^2 {\bm k} \over (2\pi)^2} (\tilde{\varepsilon}_{-,{\bm k}} - \varepsilon_{-,{\bm k}}) \cos[k \cdot ({\bm r}_j - {\bm r}_i)],
\end{align}
and
\begin{equation}\label{eq:mu}
\delta \mu = \int{d^2 {\bm k} \over (2\pi)^2} (\tilde{\varepsilon}_{-,{\bm k}} - \varepsilon_{-,{\bm k}}).
\end{equation}
The operators $\beta_{i,-}$ and $\beta_{i,-}^\dagger$ can then be understood as renormalized TMBS annihilation and creation operators, which hybridize with the excitations of two single magnons at ${\cal O}(1/M)$; here we neglect this hybridization, as it generates effects at orders higher than ${\cal O}(1/M)$. 
Including TMBS renormalization in the original spin Hamiltonian and replacing $\beta_{i,-}$ and $\beta_{i,-}^\dagger$ by the spin-1 operators in $\tilde{H}_{2,{\rm TMBS}}$ yields the full renormalized Hamiltonian capable of describing the FQ phase (and hence the FP-FQ transition) in the form
\begin{align}\label{eq:Hfq}
{\cal H}_\text{FQ} = {\cal H} + \tfrac{1}{2} \delta t_{ij} \left( Q_{i}^{+} Q_{j}^{-} + {\rm h.c.} \right)  - \tfrac{1}{2} {\delta \mu} \sum_{i} \left( S_{i}^{z} \right)^{2} + \tfrac{1}{2} {\delta\mu} \sum_{i} S_{i}^{z},
\end{align}
where $Q_{i}^{\pm} = (S_{i}^{\pm})^{2}$. This expression was used to compute the dynamical structure factor shown in Fig.~4A1 of the main text.

\begin{figure}[h!] 
 	\begin{center}
 		\includegraphics[width=0.9\textwidth]{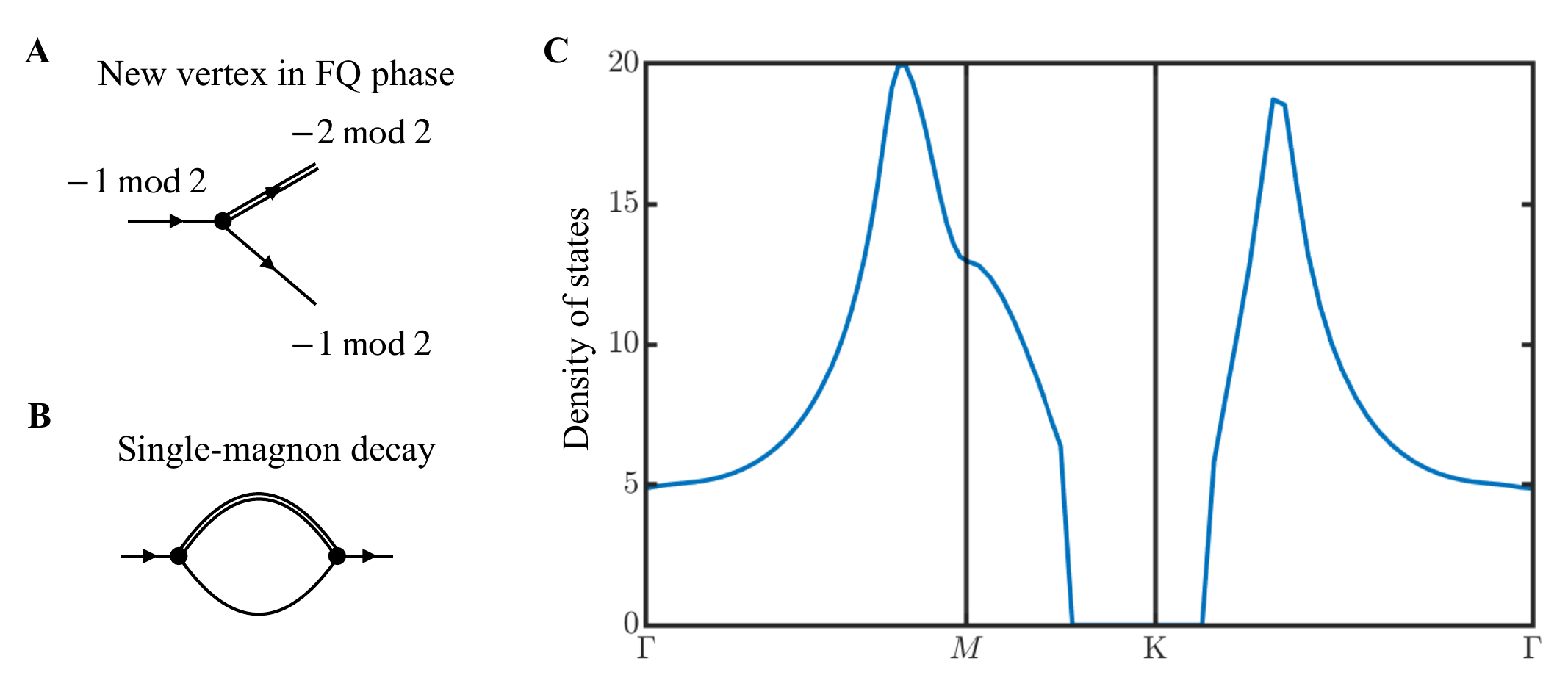}
	\end{center}
\caption{\label{SI_decay} ({\bf A}) Cubic vertex when ferroquadrupolar ordering develops in the FQ phase. As in Fig.~S\ref{SI_vertex}, numbers label the spin parity associated with each propagator. ({\bf B}) Leading-order single-magnon decay process in the FQ phase, mediated by a TMBS. ({\bf C}) Density of states of decay products described in panel (B), which include a TMBS (the Goldstone mode) and a single-magnon excitation.}
\end{figure}

The renormalized Hamiltonian ${\cal H}_\text{FQ}$ [Eq.~\eqref{eq:Hfq}] describes the effective quadrupolar interaction that develops at the TMBS energy scale, which drives the FP-FQ phase transition at the classical level. It also explains the universal dynamics of the pair condensate in the FQ phase, where the ${\mathrm U}(1) \to {\mathbb Z}_2$ symmetry-breaking makes the spin parity a conserved quantity but rather than the total spin. The new cubic vertex appearing as a consequence, illustrated in Fig.~S\ref{SI_decay}A, introduces a new decay channel for single magnons, represented by the diagram in Fig.~S\ref{SI_decay}B. Evaluated on-shell (i.e.~for an external frequency equal to the single-magnon dispersion), the imaginary part of this diagram gives the single-magnon decay rate represented by the yellow shading in Fig.~4B1 in the main text, which alongside the dispersion is one of the key signatures of universal dynamics.

In detail, this decay rate is a convolution of the absolute square of the cubic vertex with the density of states of the decay products, which are another single magnon and a TMBS (which is the Goldstone mode of the FQ phase). The momentum phase space for these decay products in momentum space is illustrated explicitly for each initial magnon wave vector ${\bm q}$ in Fig.~4B2 of the main text. The density of states along the same momentum path as in Fig.~4B1 of the main text is shown in Fig.~S\ref{SI_decay}C, and we stress that it is peaked around M and X$^\prime$ while being suppressed to zero around K, consistent with the distribution of the decay rate itself. In Fig.~2C2 of the main text, we computed the dynamical spin structure factor by retaining the full frequency-dependence of the self-energy correction, which captures both the broadening of the single-magnon band in the on-shell approximation and the spectral-weight transfer to the high-energy continuum.

In Na$_2$BaNi(PO$_4$)$_2$, it is easy to estimate $J \simeq 0.0348$~meV on the basis of the single-magnon dispersion in the FP phase [Eq.~\eqref{eq:smd}]. It is more difficult to separate the anisotropy contributions from the $D$ and $\Delta$ terms, which we defer to Sec.~S2.5. Nevertheless, for reasonable estimates of $D$ we obtain the result $h_c \simeq 1.96$~T from Eq.~\eqref{eq:hc-gswt}, which is slightly larger than both the experimental result and the result $h_c = 1.88$~T obtained from exact diagonalization, indicating the limits of the renormalized GSWT procedure at this order. For the renormalized TMBS, from Eq.~\eqref{eq:tij} we obtain $\delta t^{{\rm n.n.}} \simeq - 0.00425$~meV for the hopping correction in the situation where $i$ and $j$ are nearest-neighbor (n.n.) sites and $\delta t^{{\rm n.n.n.}} \simeq - 0.000974$~meV where they are next-nearest neighbors (all longer-range terms being much smaller); from Eq.~\eqref{eq:mu} we obtain $\delta \mu \simeq 0.0327$~meV, indicating the extent to which single-magnon dispersion and binding influence the Goldstone mode of the FQ phase in \NiTL.

\subsubsection{Renormalized Hamiltonian for UUD and NSS phases}

The UUD phase has collinear, three-sublattice magnetic order with the local moments on sublattices $1$ and $2$ oriented ``up'' while those on sublattice $3$ are aligned ``down.'' We label the lattice sites by $(i,m)$, where $i$ indexes the unit cell and $m = 1,2,3$ the sublattice within each cell. The local reference frame is defined by the transformations $\beta_{(i,m),\sigma} = b_{(i,m),\sigma}$ for $m = 1$ and $2$, while for $m = 3$ $\beta_{(i,3),+} = b_{(i,3),-}$, $\beta_{(i,3),0} = -b_{(i,3),0}$, and $\beta_{(i,3),-} = b_{(i,3),+}$. In this frame it is the condensation of $\beta_{(i,m),+}$ for all $m$ that gives UUD magnetic order, whence the GSWT Hamiltonian takes the form
\begin{eqnarray}
    {\cal H}_{2} & = & J \sum_{i,{\bm \delta}} \left[ \beta_{(i,1),0}^\dagger \beta_{(i,1)+{\bm \delta},0} - \beta_{(i,2),0}^\dagger \beta_{(i,2)+{\bm \delta},0} -
    \beta_{(i,3),0}^\dagger \beta_{(i,3)+{\bm \delta},0} + {\rm h.c.} \right] \nonumber \\
    & & \quad + \sum_i (D + h) \left( n_{(i,1),0} + n_{(i,2),0} \right) + (D - h + 6J \Delta) n_{(i,3),0} \nonumber \\
    & & \quad + \sum_i 2h \left( n_{(i,1),-} + n_{(i,2),-} \right) + \left( 12J \Delta - 2h \right) n_{(i,3),-},
\end{eqnarray}
where $(i,m) + \bm{\delta}$ labels the site $(i',m')$ connected to site $(i,m)$ by each of the three vectors ${\bm \delta}$. As noted in the main text, the U(1) spin rotation symmetry means that each single-particle eigenstate of ${\cal H}_2$ has a well-defined spin quantum number, $\Delta S^z$, measured relative to the ground state. The spectrum contains three dispersive single-magnon modes, two with $\Delta S^z = -1$ and one with $\Delta S^z = 1$, and three flat TMBS bands that can be considered as localized on each sublattice. The TMBSs on sublattices $1$ and $2$ have energy $\varepsilon_{-,1} = \varepsilon_{-,2} = 2h$ and spin quantum number $\Delta S^z = -2$ while the TMBS on sublattice $3$ has $\varepsilon_{-,3} = 12J \Delta - 2h$ and $\Delta S^z = 2$.

Again these TMBS bands become dispersive when higher-order quantum fluctuations are considered. The U(1) symmetry ensures that the normal Green functions and self-energies of the TMBSs are both block diagonal,
\begin{eqnarray}
    {\cal D}(k) = {\cal D}_{uu}(k) \otimes {\cal D}_d(k), \quad \Pi(k) = \Pi_{uu}(k) \otimes \Pi_d(k),
\end{eqnarray}
where $k \equiv (\bm{k}, \omega)$ with the momentum ${\bm k}$ spanning the reduced Brillouin zone. These Green functions and self-energies are related by the Dyson equations
\begin{eqnarray}
    i {\cal D}_{uu}(k) & = & [\varepsilon_{uu} - \omega - i \Pi_{uu}(k)]^{-1}, \quad \varepsilon_{uu} = \mathrm{diag}[\varepsilon_{-,1}, \varepsilon_{-,2}], \\
    i {\cal D}_{d}(k) & = & [\varepsilon_{d} - \omega - i \Pi_{d}(k)]^{-1}, \quad  \ \  \ \ \ \ \varepsilon_{d} = \varepsilon_{-,3}.
\end{eqnarray}
At one-loop order, the self-energy has two contributions, $\Pi(\bm{k}, \omega) = \Pi^{(1)}(\bm{k}, \omega) + \Pi^{(2)}(\bm{k})$, with the first term arising from the cubic interaction, ${\cal H}_3$, and the second from the quartic interaction, ${\cal H}_4$. 

To analyze $\Pi^{(1)}(\bm{k}, \omega)$ we express the cubic interaction 
\begin{eqnarray}
    {\cal H}_3 & = & J \sum_{i,{\bm \delta}} \Big[ \beta_{(i,1),-}^\dagger \beta_{(i,1),0} \beta_{(i,1)+{\bm \delta},0} 
    + \beta_{(i,1)+{\bm \delta},-}^\dagger \beta_{(i,1)+{\bm \delta},0} \beta_{(i,1),0} \nonumber \\
    && \quad - \beta_{(i,2),0}^\dagger \beta_{(i,2),-} \beta_{(i,2)+{\bm \delta},0} 
    - \beta_{(i,2)+{\bm \delta},0}^\dagger \beta_{(i,2)+{\bm \delta},-} \beta_{(i,2),0} \nonumber \\
    && \quad - \beta_{(i,3),0}^\dagger \beta_{(i,3),-} \beta_{(i,3)+{\bm \delta},0} 
    - \beta_{(i,3)+{\bm \delta},0}^\dagger \beta_{(i,3)+{\bm \delta},-} \beta_{(i,3),0} \Big] + \mathrm{h.c.}
\end{eqnarray}
in momentum space as
\begin{eqnarray}\label{eq:si_cubic_uud}
    {\cal H}_3 = \frac{1}{2! \sqrt{N}} \sum_{\bm{k}, \bm{q}}\sum_{m=1}^3 \sum_{n,l=1}^3\tilde{V}_{m}^{nl}(\bm{k}, \bm{q}) \beta_{\bm{k}, (m,-)}^\dagger \psi_{\bm{q}, n} \psi_{\bm{k}-\bm{q}, l} + \mathrm{h.c.},
\end{eqnarray}
where $\beta_{{\bm k},(m,\sigma)} = (1/\sqrt{N_u}) \sum_i \beta_{(i,m),\sigma} \exp(i {\bm k} \cdot {\bm r}_{(i,m)})$, with $N_u = N/3$ the number of unit cells and $\psi_{{\bm q},n}$ the $n$-component of the Nambu spinor $\psi_{\bm{k}} = (\beta_{\bm{k}, (1,0)}, \beta_{\bm{k}, (2,0)}, \beta_{\bm{k}, (3,0)}, \beta^\dagger_{-\bm{k}, (1,0)}, \beta^\dagger_{-\bm{k}, (2,0)}, \beta^\dagger_{-\bm{k}, (3,0)})^T$ describing the single-magnon modes. The symmetrized cubic interaction vertex is given by ${\tilde V}_{m}^{nl}(\bm{k}, \bm{q}) = V_{m}^{nl}(\bm{k}, \bm{q}) + V_{m}^{ln}(\bm{k}, \bm{k}-\bm{q})$ in terms of the unsymmetrized cubic vertices
\begin{eqnarray}
    V_1(\bm{k}, \bm{q}) = 
    J
    \left(
    \begin{array}{cccccc}
       0  & \gamma_{\bm{k}-\bm{q}}  &  &  &  & -\gamma_{\bm{q}-\bm{k}}  \\
        \gamma_{\bm{q}} &  0 &  &  &  &  \\
         &   & 0 &  &  &  \\
         &   &  & 0 &  &  \\
         &   &  &  & 0 &  \\
         -\gamma_{-\bm{q}} &   &  &  &  &  0\\
    \end{array}
    \right), 
    V_2(\bm{k}, \bm{q}) = 
    J
    \left(
    \begin{array}{cccccc}
       0  & \gamma_{-\bm{q}}  &  &  &  &  \\
        \gamma_{\bm{q}-{\bm k}} &  0 &  &  &  &  -\gamma_{\bm{k}-\bm{q}} \\
         &   & 0 &  &  &  \\
         &   &  & 0 &  &  \\
         &   &  &  & 0 &  \\
          & -\gamma_{\bm{q}}  &  &  &  &  0\\
    \end{array}
    \right) \nonumber
\end{eqnarray}
\begin{eqnarray}
    {\rm {and}} \;\; V_3(\bm{k}, \bm{q}) = 
    J
    \left(
    \begin{array}{cccccc}
       0  &   &  &  &  &  \\
         &  0 &  &  &  &  \\
         &   & 0 & \gamma_{\bm{k}-\bm{q}} & \gamma_{\bm{q}-\bm{k}} &  \\
         &   & \gamma_{\bm{q}} & 0 &  &  \\
         &   &  \gamma_{-\bm{q}} &  & 0 &  \\
          &   &  &  &  &  0\\
    \end{array}
    \right).
\end{eqnarray}
Using this notation, the self-energy contribution $\Pi^{(1)}_{mn}(\bm{k}, \omega)$ is expressed as
\begin{eqnarray}
    i\Pi^{(1)}_{mn}(\bm{k}, \omega) &=& \frac{1}{2} \int \frac{d^2\bm{q}}{(2\pi)^2} \sum_{a,b} \Bigg[ 
    \frac{\left( X_{\bm{k}-\bm{q}, a}^T \tilde{V}^T_m (\bm{k}, \bm{q}) X_{\bm{q}, b} \right) 
          \left( X_{\bm{q}, b}^\dagger \tilde{V}^*_n (\bm{k}, \bm{q}) X^*_{\bm{k}-\bm{q}, a} \right)}{\varepsilon_{\bm{k}-\bm{q}, a} + \varepsilon_{\bm{q}, b} - \omega} \nonumber \\
    && \quad + \left( X \to \bar{X}, \omega \to -\omega \right) \Bigg],
\end{eqnarray}
where $X_{\bm{k}, a}$ is the $a$th eigenvector of the single-magnon GSWT Hamiltonian at momentum $\bm{k}$, expressed in the Nambu basis $\psi_{\bm{k}}$.

The second contribution, $\Pi^{(2)}({\bm k},\omega)$, is the usual Hartree-Fock self-energy, which is given by
\begin{eqnarray}
    i\Pi_{uu}^{(2)}({\bm k},\omega) & = & 
    \begin{pmatrix}
       \mu_{1}  & -t_{{\bm k}} \\
         -t_{{\bm k}} & \mu_{2}
    \end{pmatrix}, \quad i\Pi_{d}^{(2)}({\bm k},\omega) = \mu_{3},
\end{eqnarray}
where
\begin{eqnarray}
    \mu_{1} & = & \mu_{2} \nonumber\\
    & = & 6\Delta J (C_2({\bm 0}) - C_3({\bm 0})) - J \sum_{{\bm \delta}} {\rm Re} \left[ C_1({\bm \delta}) - D_3({\bm \delta}) \right], \\
    \mu_{3} & = & 6 \Delta J (C_1({\bm 0}) + C_2({\bm 0})) - J \sum_{{\bm \delta}} {\rm Re} \left[ D_2({\bm \delta}) + D_3({\bm \delta}) \right], \\
    t_{{\bm k}} & = & J \sum_{{\bm \delta}} C_1({\bm \delta}) e^{-i{\bm k} \cdot {\bm \delta}} + {\rm c.c.},
\end{eqnarray}
and we have defined the quantities $C_m({\bm \delta}) = \langle \beta_{(i,m),0}^\dagger \beta_{(i,m)+{\bm \delta},0} \rangle$ and $D_m({\bm \delta}) = \langle \beta_{(i,m),0} \beta_{(i,m)+{\bm \delta},0} \rangle$.

The renormalized TMBS spectrum can be obtained in the on-shell approximation, where the self-energy is evaluated at the unrenormalized frequency, to yield the Green function
\begin{eqnarray}
    i{\cal D}({\bm k},\omega) = \left(\tilde{\cal H}_2({\bm k}) - \omega\right)^{-1},
\end{eqnarray}
where $\tilde{\cal H}_2({\bm k})$ serves as an effective Hamiltonian for the TMBS modes and has the block-diagonal form
\begin{eqnarray}
    \tilde{\cal H}_{2;uu}({\bm k}) & = & 2h - i \Pi_{uu}({\bm k}, 2h), \\
    \tilde{\cal H}_{2;d}({\bm k}) & = & 12 J \Delta - 2h - i \Pi_{d}({\bm k}, 12 J \Delta - 2h).
\end{eqnarray}
In real space, the TMBS dynamics are described by two tight-binding models. Sublattices 1 and 2 form a honeycomb lattice on which the motion of the TMBS is governed by 
\begin{eqnarray}
    \tilde{\cal H}_{2;uu} = 
    \sum_{i,j }\sum_{m,n = 1,2}  \left( \delta t_{uu}^{(i,m)(j,n)} \beta_{(j,n),-}^\dagger \beta_{(i,m),-} + \mathrm{h.c.} \right) 
    + \left( 2h - \delta \mu_{uu} \right) \sum_{i}\sum_{m = 1,2} \beta_{(i,m),-}^\dagger \beta_{(i,m),-},
\end{eqnarray}
where 
\begin{eqnarray}\label{eq:parameters_12}
    \delta \mu_{uu} & = & \frac{1}{2} \int \frac{d^2{\bm k}}{(2\pi)^2} \mathrm{Tr}\left[i \Pi_{uu}({\bm k}, 2h)\right], \\
    \delta t_{uu}^{(i,m),(j,n)} & = & \int \frac{d^2{\bm k}}{(2\pi)^2} (-i)\Pi_{uu}^{mn}({\bm k}, 2h) e^{i {\bm k} \cdot ({\bm r}_{(i,m)} - {\bm r}_{(j,n)})}.
\end{eqnarray}
Lattice sites on sublattice $3$ form a triangular lattice, on which the motion of the TMBS is described by 
\begin{eqnarray}
    \tilde{\cal H}_{2;d} = 
    \sum_{i,j } \left( \delta t_d^{(i,3)(j,3)} \beta_{(j,3),-}^\dagger \beta_{(i,3),-} + \mathrm{h.c.} \right) 
    + \left( 12 J \Delta - 2h - \delta \mu_{d} \right) \sum_{i} \beta_{(i,3),-}^\dagger \beta_{(i,3),-},
\end{eqnarray}
with
\begin{eqnarray}\label{eq:parameters_3}
    \delta \mu_{d} & = & \int \frac{d^2{\bm k}}{(2\pi)^2} i \Pi_{d}({\bm k}, 12J\Delta - 2h), \\
    \delta t_d^{(i,3)(j,3)} 
    & = & \int \frac{d^2{\bm k}}{(2\pi)^2} \, -i \Pi_{d}({\bm k}, 12J\Delta - 2h) e^{i {\bm k} \cdot ({\bm r}_{(i,3)} - {\bm r}_{(j,3)})}.
\end{eqnarray}

As for the FP and FQ phases, we estimate these effective parameters for \NiTL. Within the UUD phase they are independent of the field strength, with values $\delta \mu_{uu} \simeq 0.0137$~meV, $\delta t_{uu}^{\mathrm{n.n.}} \simeq -0.00575$~meV, and $\delta t_{uu}^{\mathrm{n.n.n.}} \simeq -0.00056$~meV for the U sublattices and $\delta \mu_{d} \simeq 0.0022$~meV and $\delta t_d^{{\rm n.n.}} \simeq -0.00006$~meV on the D sublattice. The global minimum of the three renormalized TMBS bands again occurs at the $\Gamma$ point, where field-induced TMBS condensation leads to a quadrupolar-ordered component while preserving the three-sublattice structure.


\subsection{S2.3   Exact Diagonalization Calculations}

\begin{figure}[h!] 
 	\begin{center}
 		\includegraphics[width=0.9\textwidth]{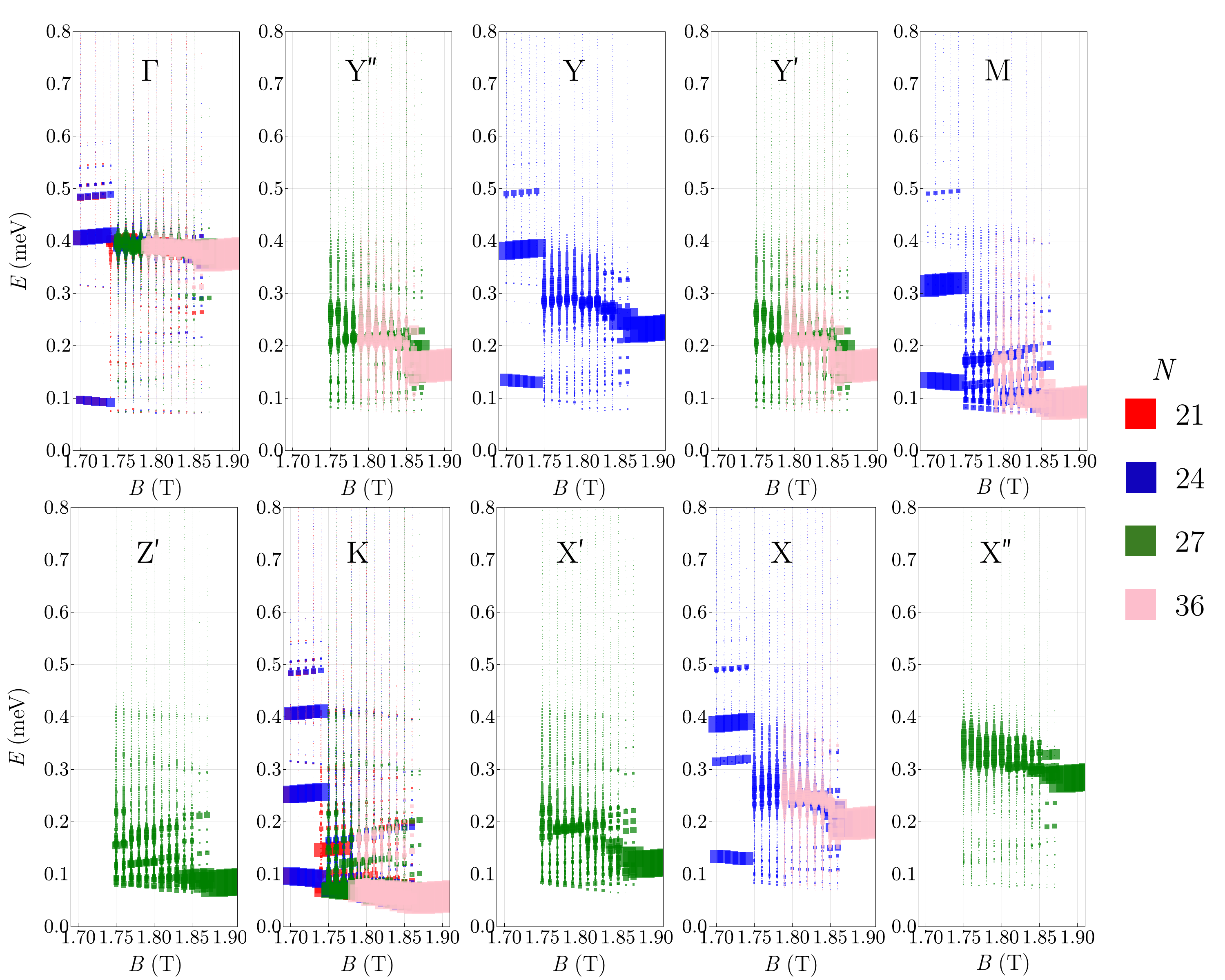}
	\end{center}
\caption{\label{poles_Sxx_HF} Full spectral data calculated by ED for the $S^{xx}$ component at magnetic fields spanning the UUD, FQ, and FP phases, combining results obtained for all cluster sizes (accessible {\bf Q} points for each cluster are shown in Fig.~S\ref{SI_crystal}B). The spectral weight of each pole is represented by the relative area of its symbol.}
\end{figure}

\begin{figure}[h!] 
 	\begin{center}
 		\includegraphics[width=0.9\textwidth]{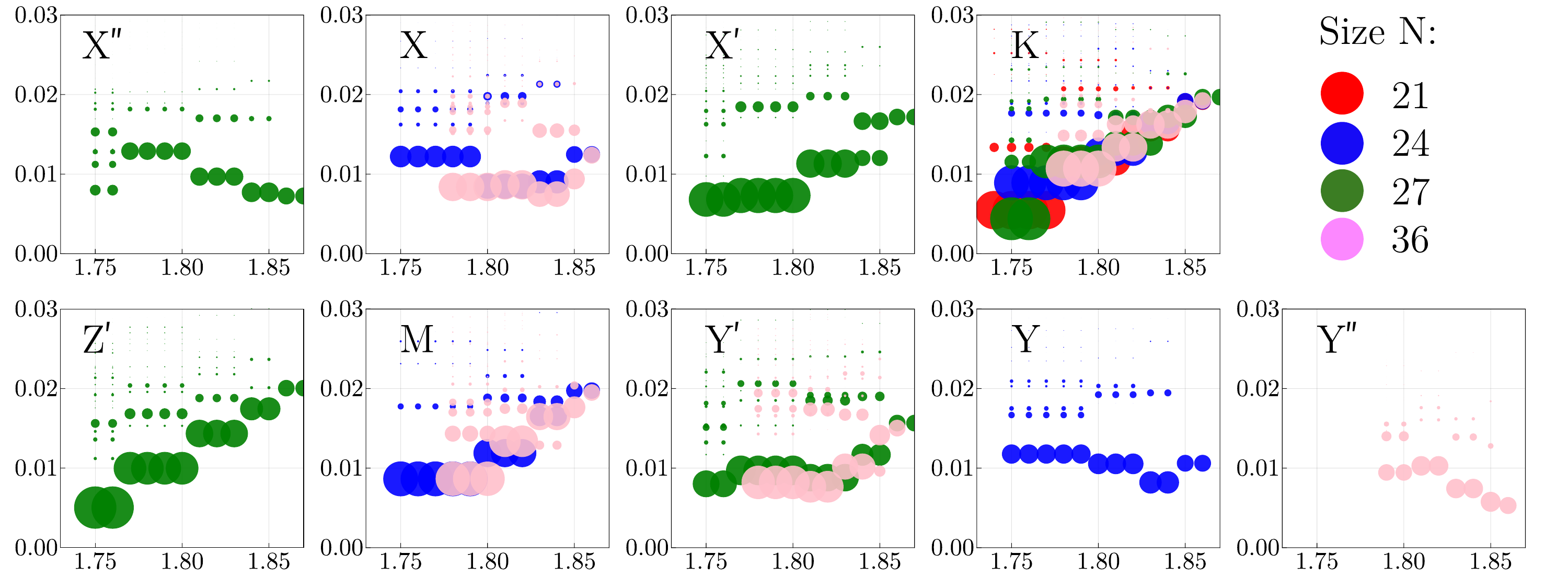}
	\end{center}
\caption{\label{poles_Szz_HF} Full spectral data calculated by ED for the $S^{zz}$ component at magnetic fields spanning the UUD, FQ, and FP phases, combining results obtained for all cluster sizes (accessible {\bf Q} points for each cluster are shown in Fig.~S\ref{SI_crystal}B). The spectral weight of each pole is represented by the relative area of its symbol.}
\end{figure}

\begin{figure}[h!] 
 	\begin{center}
 		\includegraphics[width=0.9\textwidth]{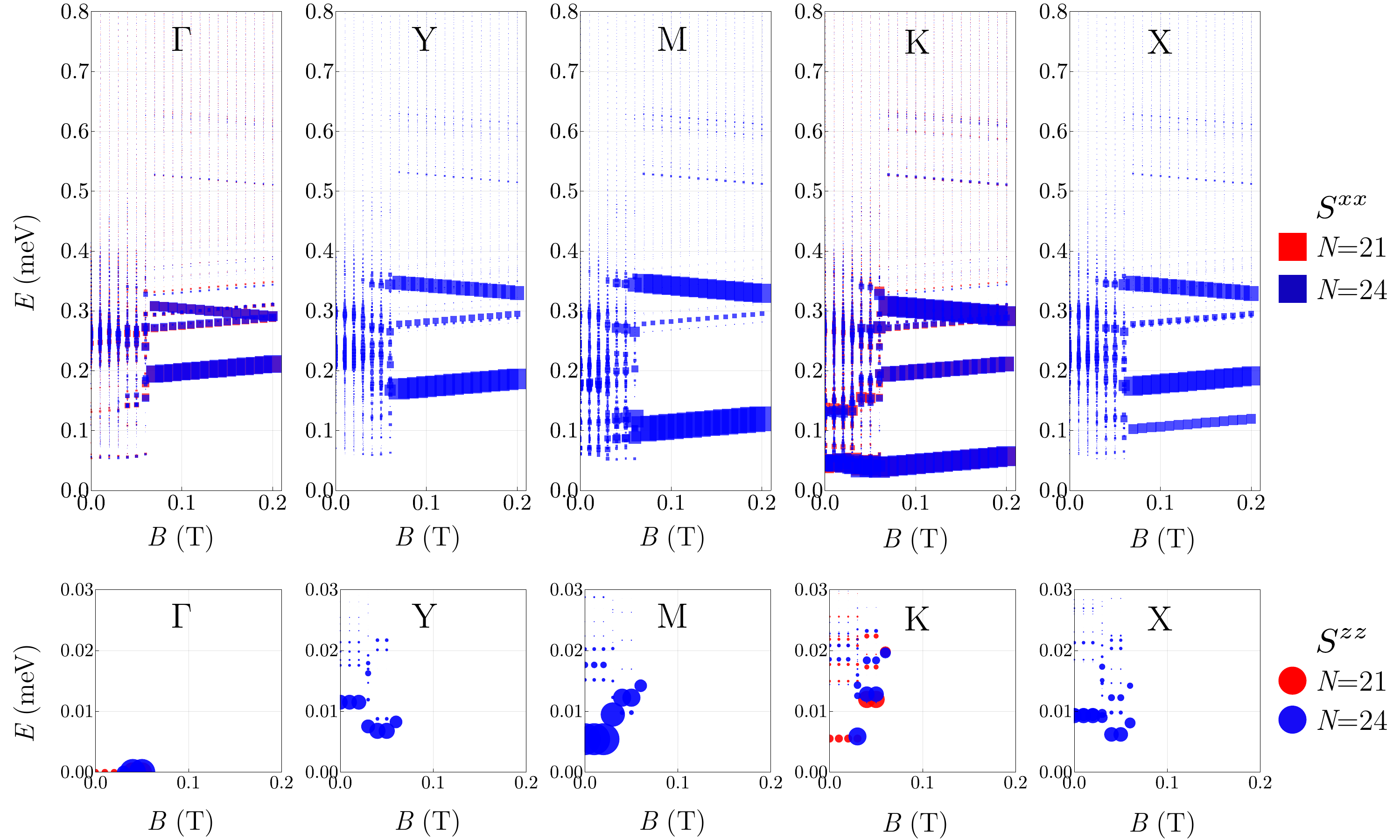}
	\end{center}
\caption{\label{poles_LF} Full spectral data calculated by ED for the $S^{xx}$ and $S^{zz}$ component at magnetic fields spanning the NSS and UUD phases, combining results obtained for all cluster sizes (accessible {\bf Q} points for each cluster are shown in Fig.~S\ref{SI_crystal}B). The spectral weight of each pole is represented by the relative area of its symbol.}
\end{figure}

As summarized in Sec.~S1.3, we compute the quantities $S^{xx}(\bQ,E)$ and $S^{zz}(\bQ,E)$ by exact diagonalization on clusters of sizes up to $N = 36$ lattice sites. The Krylov-based spectral-function calculations involve approximately 500 Hamiltonian iterations and can be evaluated in a Lehmann representation as finite sets of poles with corresponding energies and intensities. For clarity of explanation, we first present pole-based data in Figs.~\ref{poles_Sxx_HF} ($S^{xx}$) and \ref{poles_Szz_HF} ($S^{zz}$) for the high-field region containing (from left to right) the UUD, FQ, and FP phases, and in Fig.~\ref{poles_LF} for the low-field region containing the NSS and UUD phases. First, the numbers, positions, and weights of these poles show very clearly where in energy and field one may observe sharp quasiparticle peaks and where the spectrum tends towards a broader continuum of states. Second, the matrix sizes to diagonalize become larger as the total $S^z$ (and hence the effective magnetic field we apply) is reduced, until we are unable to proceed to lower fields beyond matrix sizes $M$$\times$$M$ once $M$ exceeds 8.2$\times$10$^8$. For our larger cluster sizes, the data therefore terminate at fields in the FQ phase (at 1.78~T for $N = 36$ and 1.75~T for $N = 27$, shown in Figs.~\ref{poles_Sxx_HF} and \ref{poles_Szz_HF}), whereas with $N = 24$ and below we are able to proceed to zero field (Fig.~\ref{poles_LF}). 

In \NiTL, the single magnon is strongly transverse and the Goldstone mode largely longitudinal, as a result of which the ability of ED to separate the scattering contributions $S^{xx}(\bQ,E)$ and $S^{zz}(\bQ,E)$ of Eq.~\eqref{eq:ins} is extremely valuable. Figure \ref{poles_Sxx_HF} shows most clearly how the sharp single-magnon mode of the FP phase is transformed to a broad spectrum of excitations, extending to significant energies, in the FQ phase. Figure \ref{poles_Szz_HF} shows most clearly how the dispersive Goldstone mode appears at a finite but extremely low energy in the FQ phase, with its band width being compressed further as the field is reduced towards the UUD transition. Figure \ref{poles_LF} shows that the more complex pole structure beyond the lowest magnon modes persists in the UUD phase, and that these magnons once again lose their identity, in tandem with the appearance of a low-lying Goldstone mode, below the NSS transition. 

\begin{figure}[h!] 
 	\begin{center}
 		\includegraphics[width=0.8\textwidth]{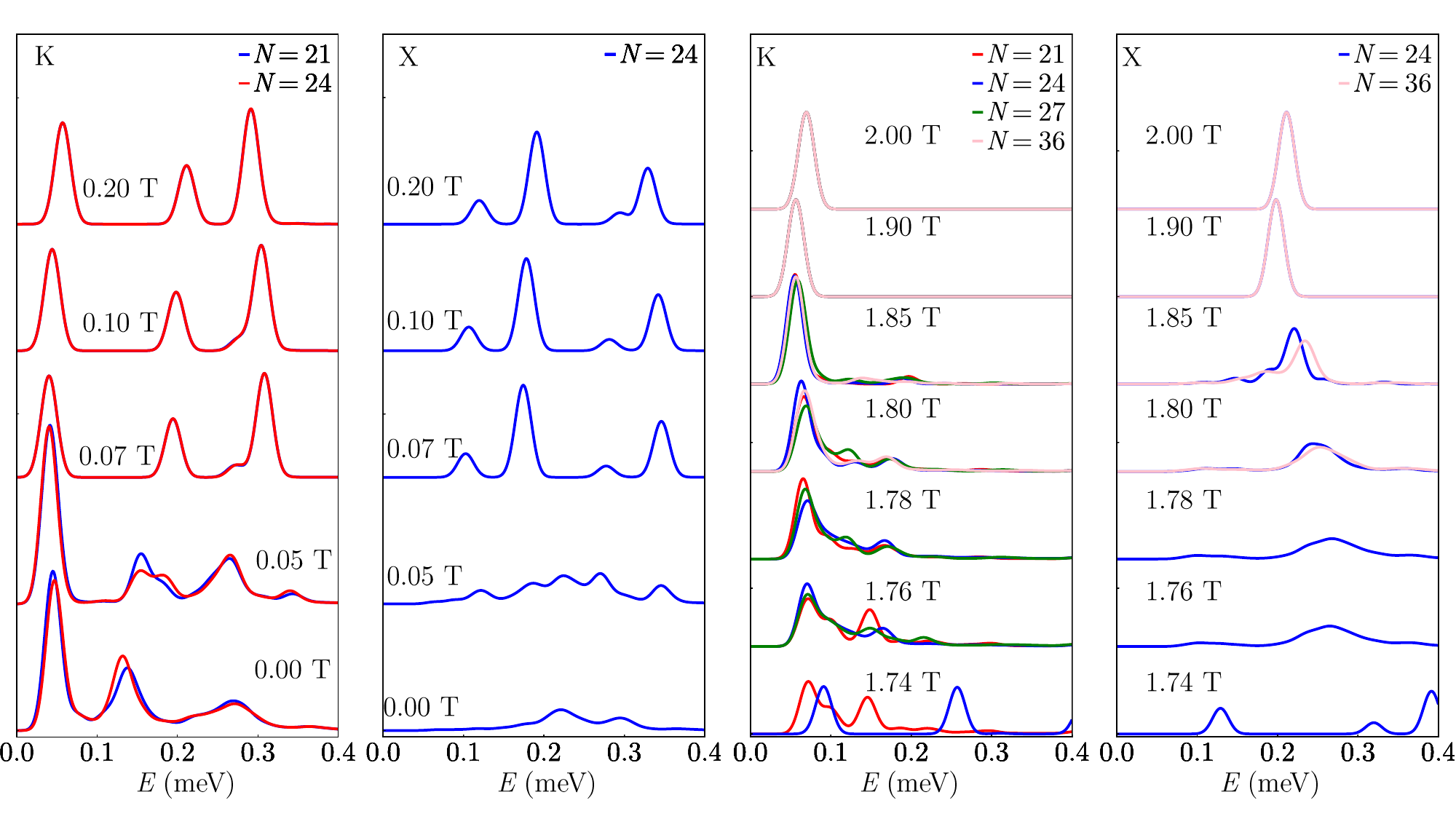}
	\end{center}
\caption{\label{ED1} $S^{xx}$ component of the spectral response computed by ED at the K and X points using the cluster sizes accessible for fields placing the system in its NSS and FQ phases.}
\end{figure}

\begin{figure}[h!] 
 	\begin{center}
 		\includegraphics[width=0.8\textwidth]{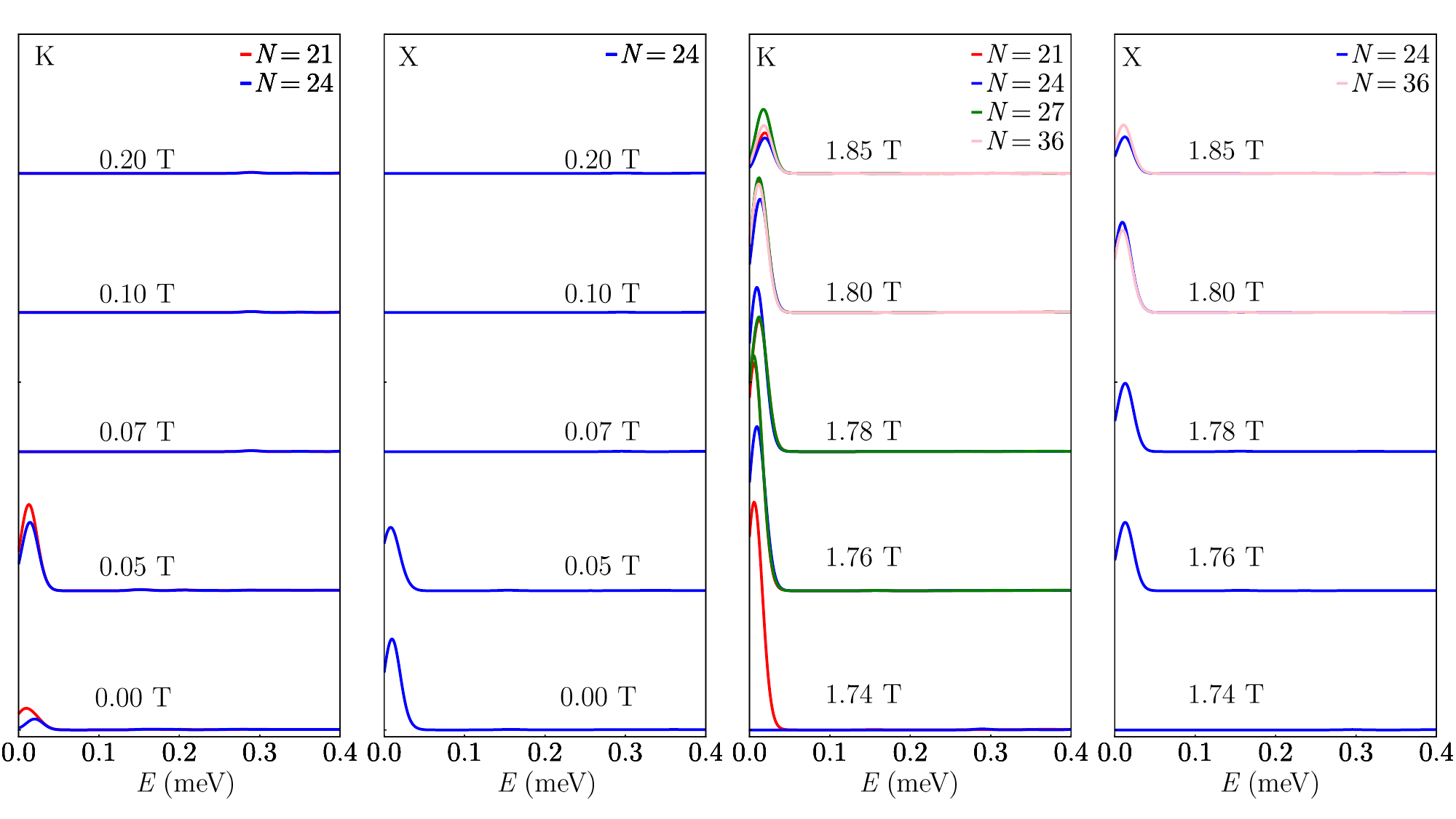}
	\end{center}
\caption{\label{ED2} $S^{zz}$ component of the spectral response computed by ED at the K and X points using the cluster sizes accessible for fields placing the system in its NSS and FQ phases.}
\end{figure}

\begin{figure}[h!] 
 	\begin{center}
 		\includegraphics[width=0.5\textwidth]{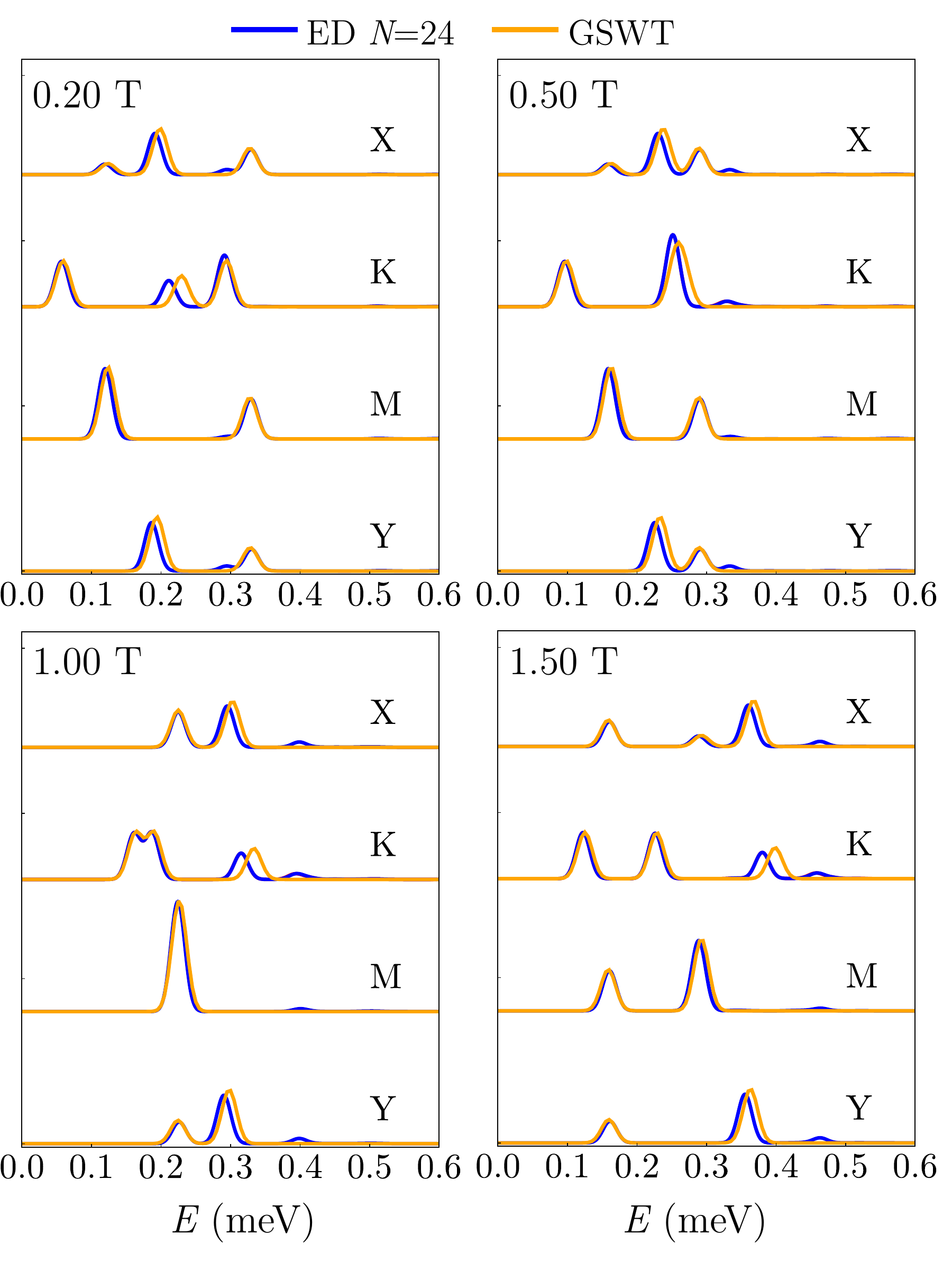}
	\end{center}
\caption{\label{ED_vs_LSW} Comparison of the $S^{xx}$ component calculated by ED and GSWT. The discrepancy originates from quantum renormalization effects that are not captured by the (linear-order) GSWT calculations, highlighting the importance of using ED to fit the experimental spectra in the UUD phase.}
\end{figure}

\begin{figure}[h!] 
 	\begin{center}
 		\includegraphics[width=0.4\textwidth]{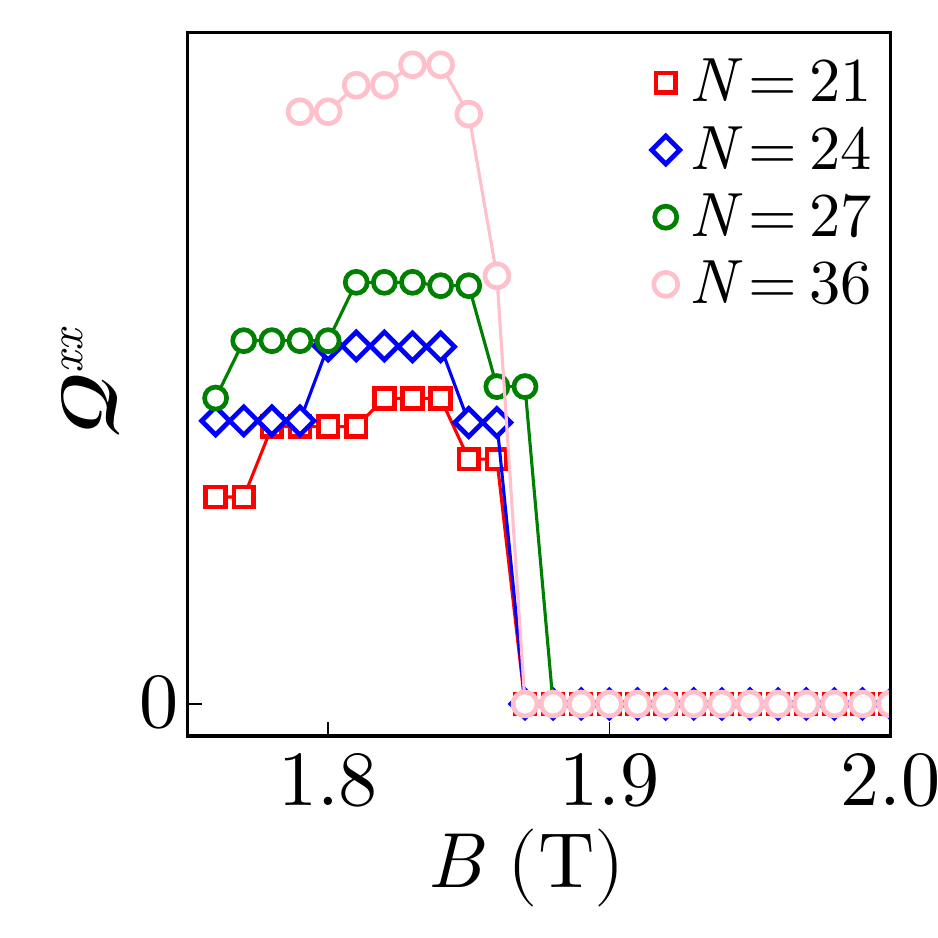}
	\end{center}
\caption{\label{ED_Qxx} Quadrupole-quadrupole correlations calculated as a function of the applied field around the FP-FQ phase transition using ED with a range of cluster sizes. }
\end{figure}

As an alternative visualization, the pole data can be shown as a continuous function of energy after applying a Gaussian broadening to mimic a finite energy resolution, which in the present work we take to be the experimental resolution ($w = 0.025$ meV FWHM). In Figs.~S\ref{ED1} ($S^{xx}$) and S\ref{ED2} ($S^{zz}$), we use this technique to display the effect of the cluster size in a number of calculations performed over the two field ranges spanning the NSS and FQ phases. This was also the way in which we displayed the ED data in Figs.~2D and 5A of the main text, while the prospects of a fit to Fig.~3 are clear in Fig.~S\ref{ED1}. In Fig.~S\ref{ED_vs_LSW}, we use our ED data to benchmark the GSWT spectra calculated in the low-field regime (NSS and UUD phases) where the lattice-translation symmetry is lowered (final subsection of Sec.~S2.2). Finally, in Fig.~S\ref{ED_Qxx} we show the quadrupole moment ($Q^{xx}$) calculated throughout the FQ phase, using a range of cluster sizes to provide perspective for the $N = 27$ result shown in Fig.~5A3 of the main text. 

\clearpage

\subsection{S2.4   Comparison between INS and ED Data}

\begin{figure}[h!] 
 	\begin{center}
 		\includegraphics[width=0.5\textwidth]{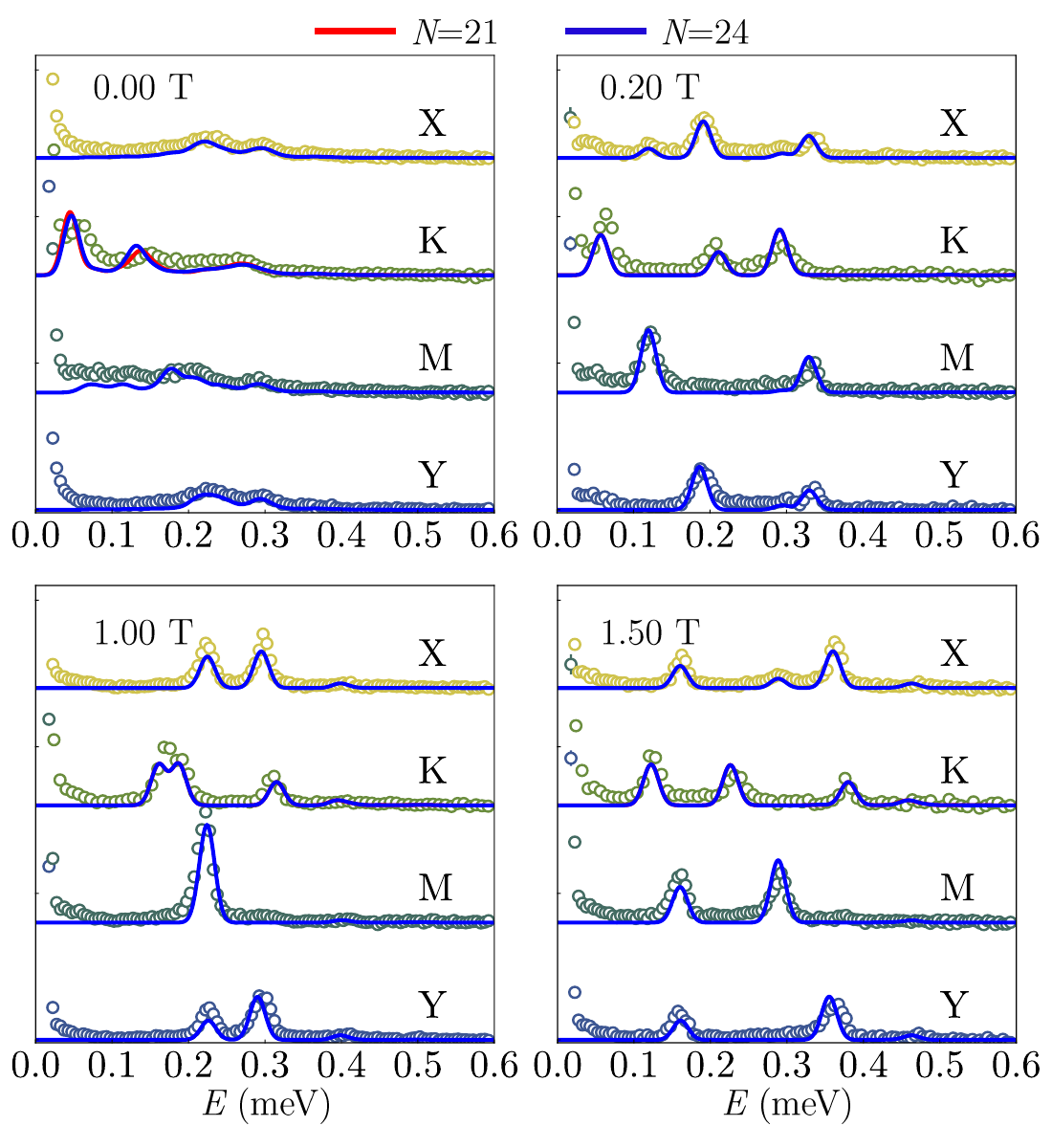}
	\end{center}
\caption{\label{INS-ED2} Comparison of INS measurements with the $S^{xx}$ component of ED data obtained at the X, K, M, and Y points using two cluster sizes for the zero-field NSS phase and three fields in the UUD phase ($B = 0.2$, $1.0$, and $1.5$~T).}
\end{figure}

\begin{figure}[h!] 
 	\begin{center}
 		\includegraphics[width=0.7\textwidth]{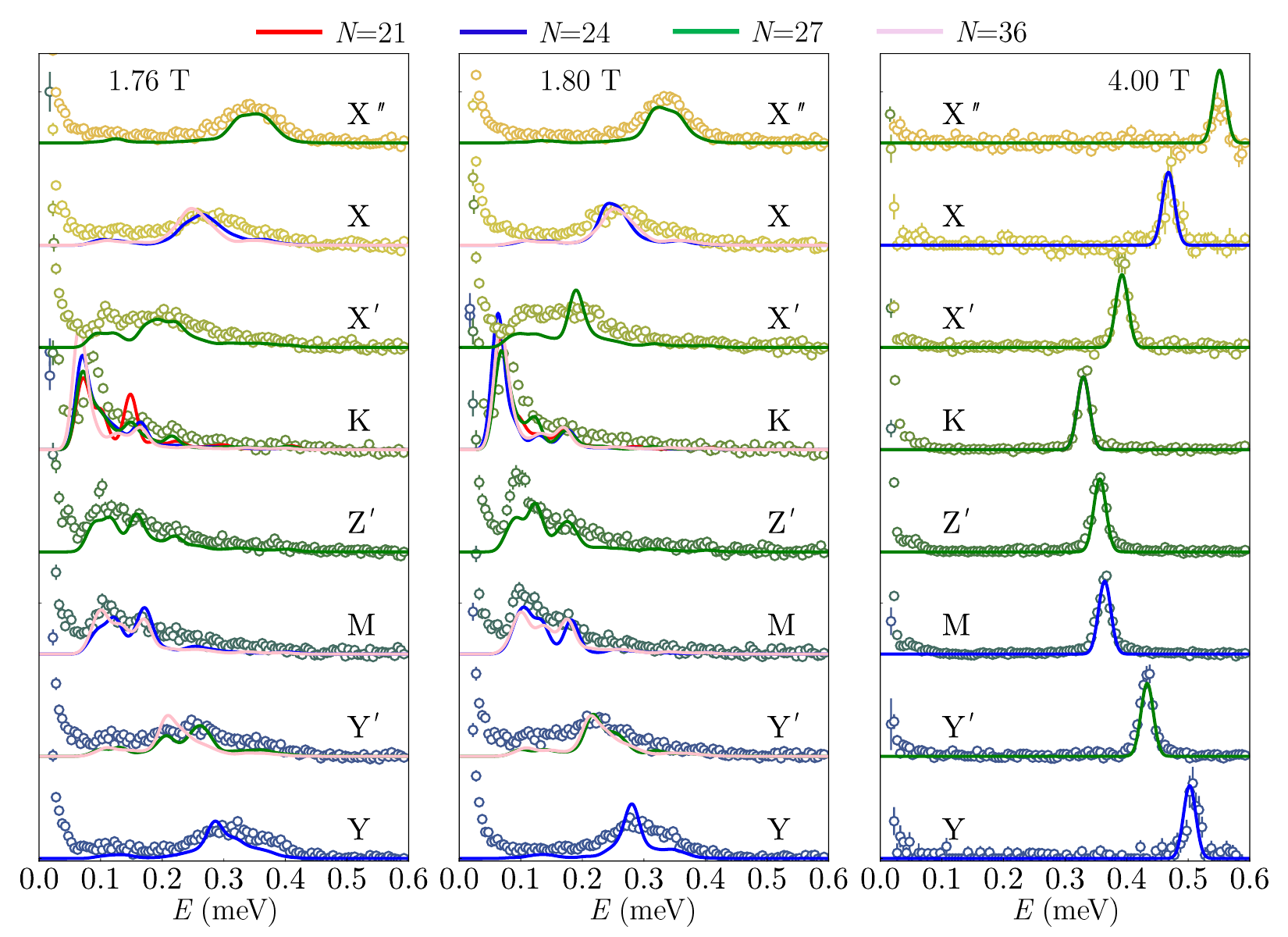}
	\end{center}
\caption{\label{INS-ED3} Comparison of INS measurements with the $S^{xx}$ component of ED data obtained at all of the high-symmetry {\bf Q} points accessible with four different cluster sizes for two fields in the FQ phase ($B = 1.76$ and $1.80$~T) and one in the FP phase ($B = 4.00$~T).}
\end{figure}

\begin{figure}[h!] 
 	\begin{center}
 		\includegraphics[width=0.9\textwidth]{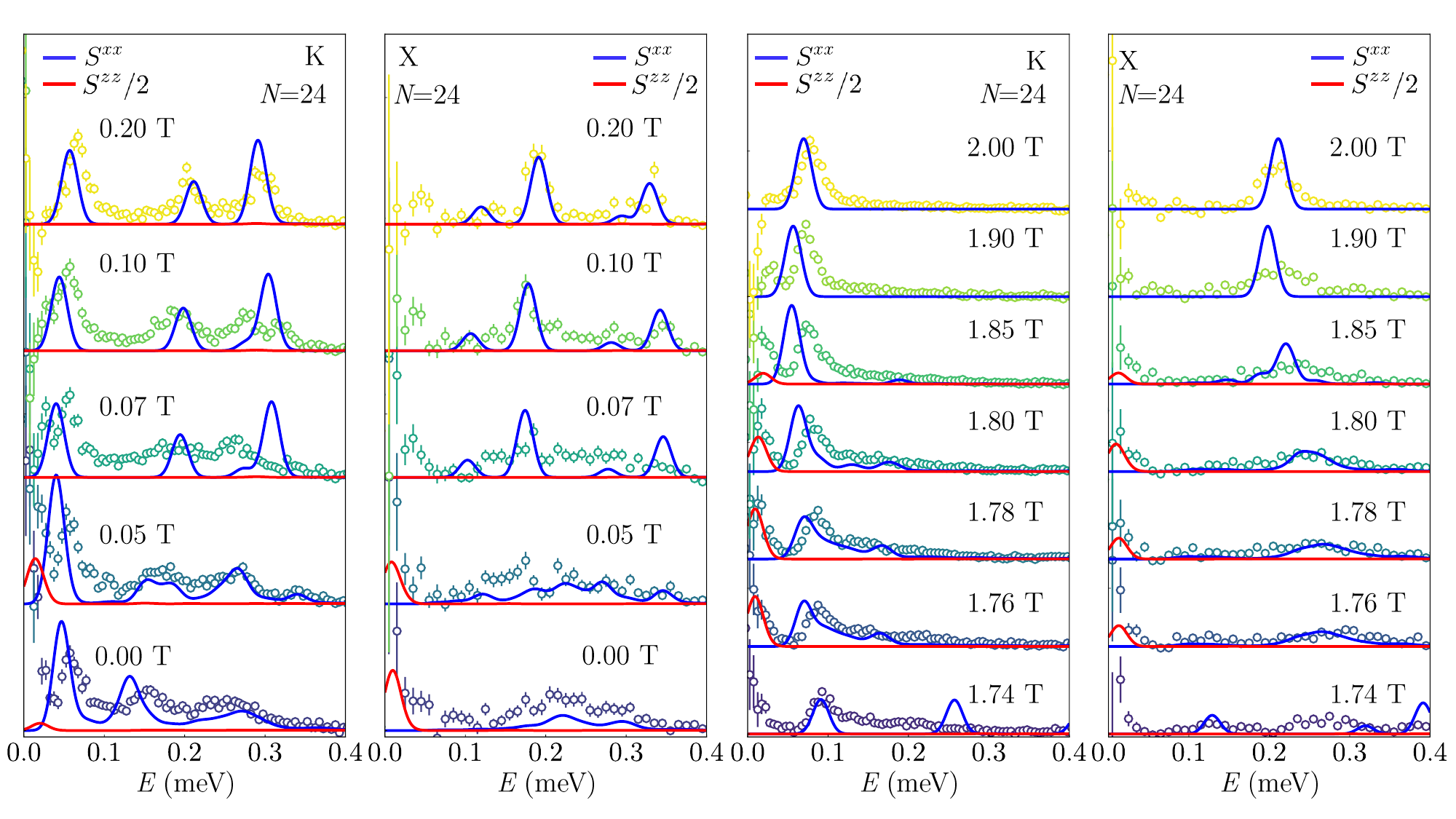}
	\end{center}
\caption{\label{INS-ED1} Comparison of INS measurements with ED data obtained for system size $N = 24$ at the K and X points for a range of fields from the UUD to the NSS phase and from the FP to the FQ phase.}
\end{figure}

Here we illustrate the quantitative comparison between our INS measurements and our ED calculations of the magnetic excitation spectra over a wide range of applied fields and for multiple wave vectors throughout the Brillouin zone. In all fits shown in Figs.~S\ref{INS-ED2} and S\ref{INS-ED3}, a single scale factor was applied to match the ED data to the measured intensities. A single scale factor also links all the panels of Fig.~S\ref{INS-ED1}, where the detailed field sweeps were measured in our second INS experiment, but this factor is different from the one used in Figs.~S\ref{INS-ED2} and S\ref{INS-ED3}. These comparisons were used to reinforce the optimal fit of the Hamiltonian parameters discussed in Sec.~S2.5. 

\subsection{S2.5   Determination of Model Parameters}

\begin{figure}[h!] 
 	\begin{center}
 		\includegraphics[width=0.9\textwidth]{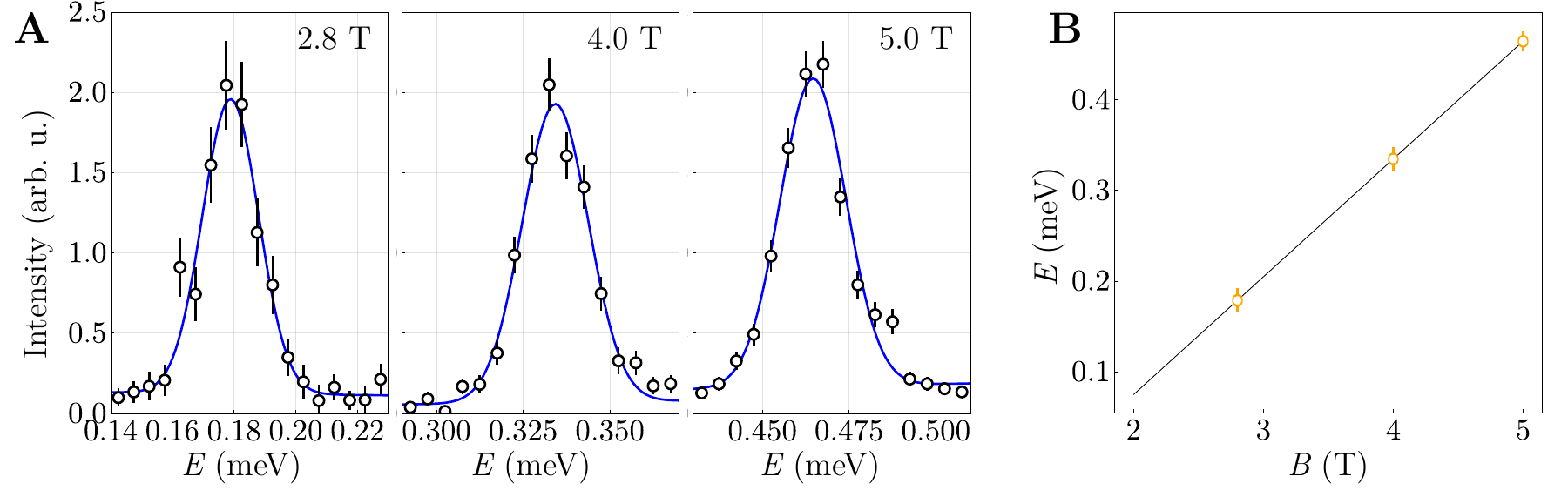}
	\end{center}
\caption{\label{SI_gfactor} ({\bf A}) Gaussian peak fits to the K-point spectrum extracted from INS data taken at $T = 0.05$~K in the FP phase with ${\textbf{B}} \parallel {\textbf{c}}$. ({\bf B}) Linear fit of the single-magnon energies at the K point fitted from panel (A). }
\end{figure}

\begin{figure}[h!] 
 	\begin{center}
 		\includegraphics[width=0.9\textwidth]{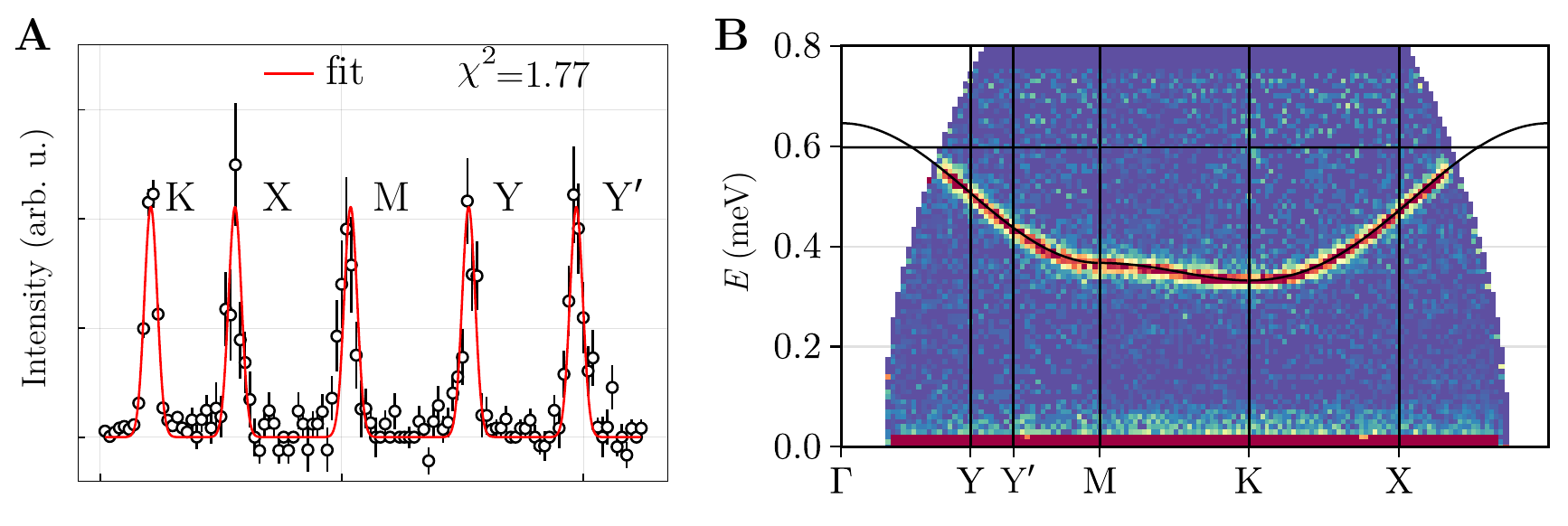}
	\end{center}
\caption{\label{SI_fits} 
 ({\bf A}) Parameter fitting to the model given in Eq.~(1) of the main text, based on energy cuts at the K, X, M, Y, and Y$'$ points. The INS data were collected at $T = 0.05$~K in the FP phase ($\textbf{B} = 4$~T $\parallel$ $\textbf{c}$). 
 ({\bf B}) Comparison of the INS spectrum along the high-symmetry path at 4~T with GSWT modelling (black lines) using the parameters of Table S\ref{table:fit}.}
\end{figure}

The minimal Hamiltonian that describes the physics of \NiTL, given in Eq.~(1) of the main text, has four unknown parameters. The $g$-factor was determined by linear fitting of the single-magnon energies at the K point over a range of applied magnetic fields in the FP phase. As shown in Fig.~S\ref{SI_gfactor}, fitting the data obtained with $B = 2.8$, 4, and 5~T yields a $g$-factor of 2.245(6). To determine the three parameters governing the dispersion, we first fit the single-magnon branch in the FP phase along the high-symmetry path shown in Fig.~S\ref{SI_fits} using GSWT, which is exact in this phase (Sec.~S2.2). While this determines $J$ uniquely and accurately, it does not separate the XXZ anisotropy $\Delta$ and the single-ion anisotropy $D$, because these enter the dispersion in the single, accurately determined combination $- 6 \Delta J + D$. To separate the two anisotropy  contributions, we extend the fitting procedure to include our spectra taken in the UUD phase, where our ED calculations provide an accurate fit. The long-ranged antiferromagnetic order of the UUD phase acts to suppress quantum fluctuations, making finite-size effects small in comparison with the FQ phase. Here our ED results are preferable to our GSWT results in accounting correctly for the quantum renormalization of single-magnon bands, as we showed with the comparison in Fig.~S\ref{ED_vs_LSW}. The best fit of the single-magnon energies at the K point to our ED results for a range of different fields gave the definitive parameters shown in Table S\ref{table:fit}. These best-fit parameters were then used for the comparisons of our INS data and ED results shown for a wide range of fields (phases) and momentum points in Figs.~S\ref{INS-ED2} to S\ref{INS-ED1}.

\clearpage

\begin{table}[h]
	\caption{Best-fit parameters. $J$, $D$, and $w$ are measured in meV while $\Delta$, $g$, and $\chi^2$ are dimensionless. } 
	\centering 
        \setlength{\tabcolsep}{10pt} 
        \renewcommand{\arraystretch}{1}
	\begin{tabular}{c  c  c  c  c || c  c } 
		\hline\hline 
		$J$ & $- 6 \Delta J + D$ & $\Delta$ & $D$ & $g$ & $w$ & $\chi^2$ \\ [0.5ex] 
		\hline 
		0.0348(2) & $-0.0834(5)$ & 1.06(1) & 0.138(1) & 2.245(6) & 0.025(1) & 1.77 \\ 
		\hline 
	\end{tabular}
	\label{table:fit} 
\end{table}

\end{document}